\documentclass[iop]{emulateapj}

\def \solar{_\odot}
\def \pow10#1{\times 10^{#1}}

\tighten

\begin{document}

\submitted{ApJ accepted version}

\title{Mass growth and mergers: direct observations of the luminosity
  function of LRG satellite galaxies out to $\MakeLowercase{z}=0.7$
  from SDSS and BOSS images}

\author{Tomer Tal\altaffilmark{1},
  David A. Wake\altaffilmark{1},
  Pieter G. van Dokkum\altaffilmark{1},
  Frank C. van den Bosch\altaffilmark{1},
  Donald P. Schneider\altaffilmark{2},
  Jon Brinkmann\altaffilmark{3} and
  Benjamin A. Weaver\altaffilmark{4}}
\altaffiltext{1}{Yale University Astronomy Department, P.O. Box
  208101, New Haven, CT 06520-8101 USA}
\altaffiltext{2}{Department of Astronomy and Astrophysics, The
  Pennsylvania State University, 525 Davey Laboratory, University
  Park, PA 16802, USA}
\altaffiltext{3}{Apache Point Observatory, Apache Point Road, P.O. Box
  59, Sunspot, NM 88349, USA}
\altaffiltext{4}{Center for Cosmology and Particle Physics, New York
  University, New York, NY 10003, USA}

\shorttitle{The luminosity function of LRG satellite galaxies}
\shortauthors{Tal et al.}

\begin{abstract}
  We present a statistical study of the luminosity functions of
  galaxies surrounding luminous red galaxies (LRGs) at average
  redshifts $\langle z\rangle=0.34$ and $\langle z\rangle=0.65$.
  The luminosity functions are derived by extracting source
  photometry around more than 40,000 LRGs and subtracting foreground
  and background contamination using randomly selected control
  fields.
  We show that at both studied redshifts the average luminosity
  functions of the LRGs and their satellite galaxies are poorly fitted
  by a Schechter function due to a luminosity gap between the centrals
  and their most luminous satellites.
  We utilize a two-component fit of a Schechter function plus a
  log-normal distribution to demonstrate that LRGs are typically
  brighter than their most luminous satellite by roughly 1.3
  magnitudes.
  This luminosity gap implies that interactions within LRG
  environments are typically restricted to minor mergers with mass
  ratios of 1:4 or lower.
  The luminosity functions further imply that roughly 35\% of the mass
  in the environment is locked in the LRG itself, supporting the idea
  that mass growth through major mergers within the environment is
  unlikely.
  Lastly, we show that the luminosity gap may be at least partially
  explained by the selection of LRGs as the gap can be reproduced by
  sparsely sampling a Schechter function.
  In that case LRGs may represent only a small fraction of central
  galaxies in similar mass halos.
\end{abstract}


\section{\label{intro}Introduction}
 Massive galaxies in the nearby universe typically have very
 little cold gas, they host old stellar populations and exhibit
 extremely low specific star formation rates
 \citep[e.g.,][]{faber_variations_1973, peletier_elliptical_1989,
   worthey_mg_1992, jrgensen_e_1999, trager_stellar_2000,
   kauffmann_stellar_2003, balogh_bimodal_2004, hogg_dependence_2004,
   thomas_epochs_2005}.
 Therefore, studies of these galaxies typically find that essentially
 all of the stellar mass growth takes place through mergers and
 other gravitational interactions, with the relative importance of
 each process still debatable.
 For example, while some authors find that major dry mergers
 contribute significantly to the mass evolution of massive galaxies
 \citep[e.g.,][]{van_dokkum_high_1999, tran_spectroscopic_2005,
   van_dokkum_recent_2005, bell_merger_2006, boylan-kolchin_red_2006,
   naab_properties_2006, mcintosh_ongoing_2008, bundy_greater_2009}, 
 others find only a mild contribution or none at all
 \citep[e.g.,][]{patton_dynamically_2002, bundy_mass_2006,
 masjedi_very_2006, wake_2df_2006, wake_2df-sdss_2008,
 masjedi_growth_2008, tojeiro_stellar_2011}. 
 Other studies argue that minor mergers and low mass accretion events
 contribute at least some of the stellar mass growth in massive
 galaxies over a longer timescale
 \citep[e.g.,][]{kormendy_surface_1989, schweizer_correlations_1992,
   van_dokkum_recent_2005, naab_formation_2007,
   bournaud_multiple_2007, kaviraj_uv_2008,naab_minor_2009,
   bezanson_relation_2009, tal_frequency_2009,
   ramos_almeida_optical_2011, tal_faint_2011}.
 The unifying challenge of many of these analyses is that observations
 of ongoing mergers typically do not recover information
 regarding the mass ratio, and therefore implied growth, of the
 progenitor galaxies.

 As an alternative to studying galaxy mergers many authors have turned
 to examining the environments in which these objects reside.
 For example, pair counts and clustering studies observe the
 progenitor galaxies prior to their merging.
 By studying the neighbors of massive galaxies one in effect probes
 the reservoir of objects with which mergers, and consequently mass
 growth, can occur.
 However, in studies of groups and clusters this technique typically
 relies on assigning environment membership to individual satellites
 using their inferred distances from the central galaxy
 \cite[e.g.,][]{ramella_spectroscopic_2000, christlein_galaxy_2003,
   martini_spectroscopic_2006, muzzin_spectroscopic_2009,
   wilson_spectroscopic_2009, chiboucas_keck/lris_2010,
   tanaka_spectroscopically_2010}.
 This method requires that significant telescope time is
 devoted to spectroscopic measurements of all member candidates in
 order to properly assess their line-of-sight velocity and distance
 from the central galaxy.
 As a consequence, environments that are studied in this way
 are often restricted to the low redshift universe or to small
 statistical samples for which sufficiently deep data are available
 (such as the \citealp{coil_deep2_2006} and
 \citealp{bolzonella_tracking_2010} studies).
 Other techniques, such as the widely used friends-of-friends
 algorithms introduce another complication by explicitly assuming the
 size, and therefore inferred mass, of candidate groups (most of these
 algorithms are essentially a variation of the technique proposed and
 demonstrated in \citealp{huchra_groups_1982} and
 \citealp{geller_groups_1983}).
 Newer algorithms attempt to overcome the need for this assumption by
 iteratively measuring the light distribution within a given radius
 and adjusting it using the inferred mass from a halo density model
 \citep{yang_halo-based_2005, yang_galaxy_2008}.
 
 Here we follow a different approach and detect satellite galaxies
 in a purely statistical manner, in practice measuring the average
 local overdensity of galaxies around massive red galaxies.
 This technique has been applied in studies of cluster galaxies by
 many authors utilizing various data sets
 \citep[e.g.,][]{driver_dwarf_1994, gaidos_galaxy_1997,
   lobo_environmental_1997, lumsden_edinburgh-durham_1997,
   valotto_luminosity_1997, wilson_faint_1997, fairley_galaxy_2002,
   goto_morphological_2003, lin_k-band_2004, wake_environmental_2005,
   loh_color_2008, lu_recent_2009}.
 The main advantage of any statistical study of this nature is that it
 requires a minimal set of a-priori assumptions regarding the
 properties of the studied objects.
 Its main disadvantage is that no detailed information can be
 extracted regarding any single satellite.

 Throughout the paper we adopt the following cosmological parameters:
 $\Omega_m=0.3$, $\Omega_{\Lambda}=0.7$ and $H_0=70$ km s$^{-1}$
 Mpc$^{-1}$.
 
 \begin{figure}
   \includegraphics[width=0.47\textwidth]{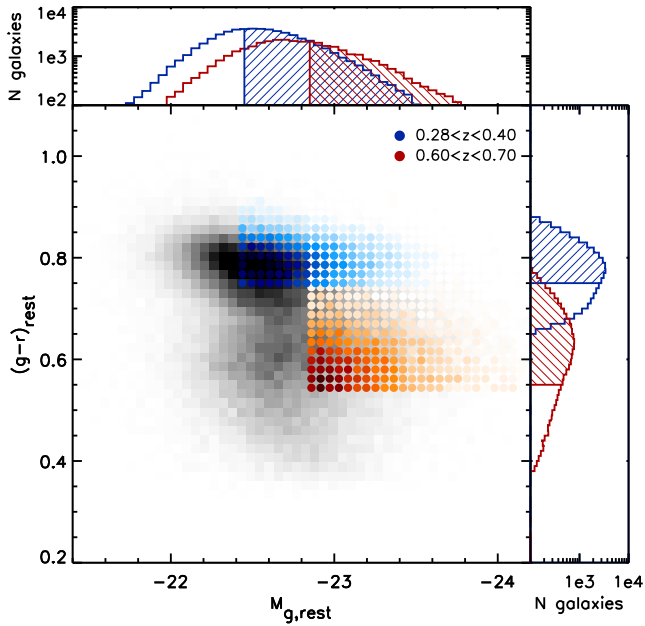}
   \caption{Rest-frame color and absolute magnitude distributions of
     the two redshift samples.
     The grayscale data points show the full SDSS/BOSS LRG data set
     within the studied redshift range while the over-plotted red and
     blue points represent the density of selected galaxies in the high 
     and low redshift bins.
     The top panel shows the initial selection of the most luminous
     galaxies at high redshift and the number density matched samples
     at low redshift.
     The right panel shows the color matching selection
     which excluded all galaxies that are bluer than the distribution
     peak value by more than 1$\sigma$.}
   \hfill
   \label{fig:colmag}
 \end{figure}
  
\section{Data}
 \subsection{Sample selection}
  We selected galaxy images for this study from the seventh data
  release of the Sloan Digital Sky Survey
  \citep[see][]{york_sloan_2000, fukugita_sloan_1996, gunn_sloan_1998,
  smith_ugriz_2002, pier_astrometric_2003, gunn_2.5_2006,
  abazajian_seventh_2009} and from the
  interim catalog of the SDSS-III Baryon Oscillation Spectroscopic
  Survey\footnote{As of January 20th 2011}
  \citep[BOSS,][]{schlegel_baryon_2009,eisenstein_sdss-iii:_2011}.
  This version of the catalog covers an area of roughly 1620 deg$^2$
  on the sky (compared to 8200 deg$^2$ in SDSS DR7) and it consists of
  more than 150,000 LRG candidates with spectra at $0.4<z<0.7$.
  All selected objects were initially identified by the SDSS pipeline
  as luminous red galaxies from their central surface
  brightness and location on a rotated color-color diagram \citep[for
    full details see][]{eisenstein_spectroscopic_2001,
    eisenstein_sdss-iii:_2011}.
  Roughly 90\% of all LRGs are central halo galaxies, thought to be
  residing in groups with a typical halo mass of a few times $10^{13}
  M_{\solar}$ \citep{wake_2df-sdss_2008, zheng_halo_2009,
    reid_constraining_2009}.
  The full data set consists of galaxies in two redshift bins as
  determined spectroscopically in the surveys.
  The low redshift sample consists of SDSS LRGs in the redshift range
  0.28${<}z{<}$0.4.
  These are the reddest, most luminous galaxies in the nearby universe
  and they occupy the high mass end of the stellar mass spectrum
  between $10^{11}M_{\solar}$ and a few times $10^{12} M_{\solar}$.
  The high redshift sample is comprised of BOSS ``CMASS'' galaxies,
  a selection aimed at finding objects at an approximately constant
  stellar mass \citep{eisenstein_sdss-iii:_2011, white_clustering_2011}.
  The color distribution of these galaxies is broader and this sample
  includes objects that are bluer than the low redshift LRGs.
  We address this in the following subsection. 
  
  \begin{figure*}
    \includegraphics[width=1.0\textwidth]{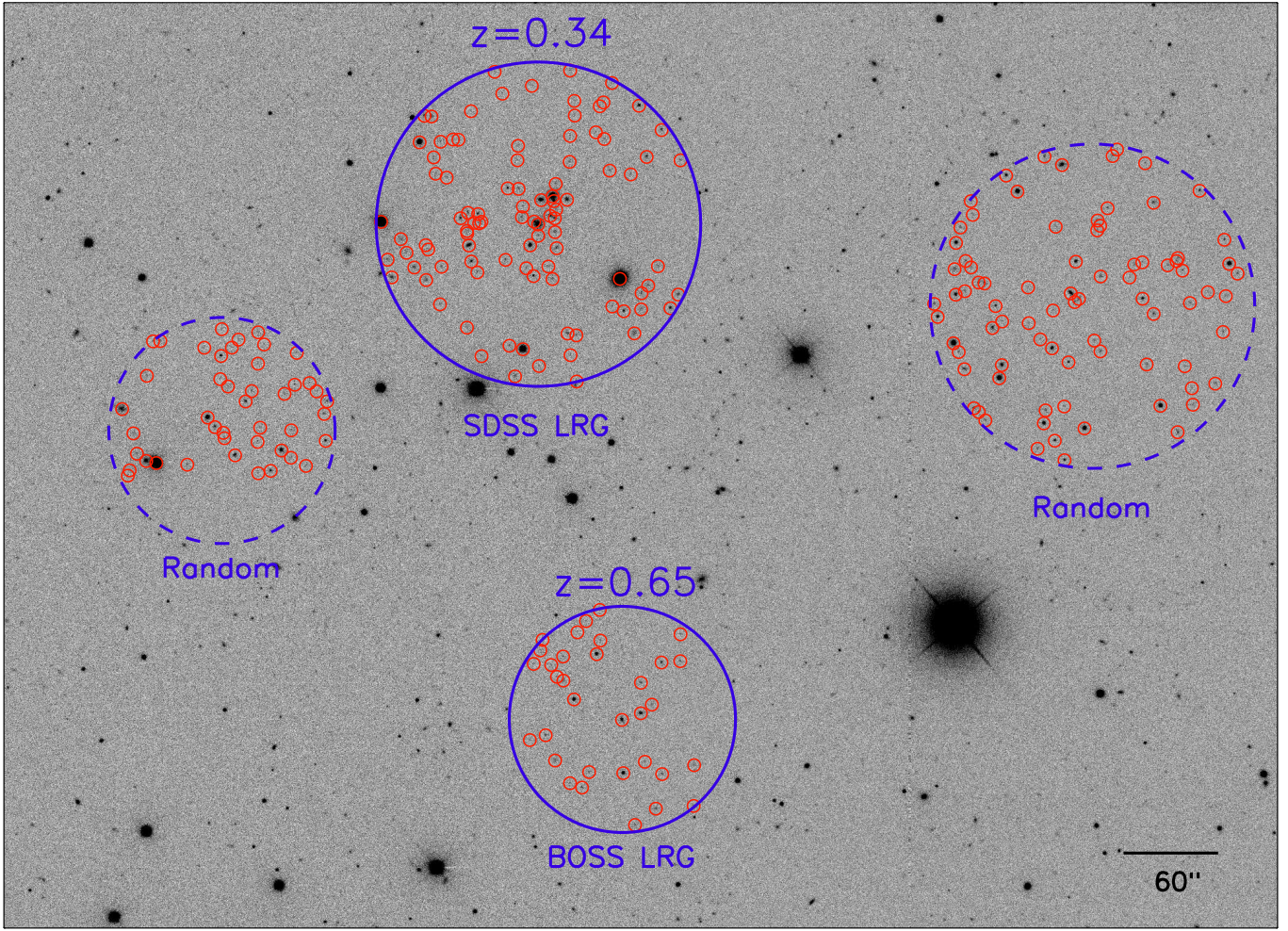}
    \caption{Examples of LRG and random aperture selections (solid and
      dashed lines, respectively) for both redshift ranges, overlaid on
      a full SDSS imaging field.
      Photometry is extracted for all objects (including LRGs) in the
      selected regions within 1000 kpc apertures using a threshold of
      1$\sigma$ above the background level.
      Objects are marked with red circles and include LRG satellites
      as well as foreground stars, foreground and background galaxies
      and spurious detections of noise peaks.
      We perform 14 and 29 random region selections and photometry
      measurements for the low- and high-redshift LRGs, respectively,
      to cover a similar area as the full imaging field.}
    \hfill
    \label{fig:lf_example}
  \end{figure*}

 \subsection{Number density and color matching} \label{ndens}
  We constructed matched samples at $z\sim 0.34$ and $z\sim 0.65$ by
  requiring that galaxies in each redshift bin have the same number
  density and a similar color distribution (with an offset due to
  passive evolution).
  This selection allows us to follow the mass evolution of massive
  galaxies, assuming that the overall number density of these objects
  does not change significantly in the studied redshift range.
  The benefits of a number-density limited, rather than luminosity- or
  mass-limited, sample have been discussed by e.g.,
  \cite{van_dokkum_growth_2010}.
  
  We started by K-correcting the SDSS extinction-corrected
  \emph{model} magnitudes in the $g-$ and $r-$bands and the BOSS
  magnitudes in the $r-$ and $i-$bands to $z=0$ following the filter
  transformation technique given by
  \cite{van_dokkum_fundamental_1996}.
  This method finds the best-fit linear combination of galaxy
  templates to an observed color and calculates the transformation
  coefficients for a given rest-frame filter.
  We used the peak of the observed color distribution as input
  ($g-r=+1.7$ and $r-i=+1.2$ for the low and high $z$ samples,
  respectively) and derived rest-frame $g-$magnitudes and $g-r$ colors
  for all galaxies in the two samples.
  The imaging bands used in this study were selected to minimize the
  necessary K-corrections as the $r-$band at $z=0.34$ and the $i-$band
  at $z=0.65$ roughly overlap with the restframe $g-$band.
  To test this, we compared the difference in derived K-correction
  values between the two redshifts with stellar population synthesis
  models \citep{bruzual_stellar_2003, maraston_modelling_2009} and
  found agreement to within a few hundredths of a magnitude. 
  
  Our initial selection included objects from the high-redshift sample
  that are brighter than $M_{g,rest}=-22.81$ mag.
  This selection minimizes the effects of incompleteness due to
  the BOSS detection limit as it excludes galaxies close to or fainter
  than the turnover point in $N(M_{g,rest})$.
  We then excluded the faintest galaxies from the low redshift sample
  until their overall number densities matched that of the high-$z$
  sample to within 1\%.
  Lastly, we fit a Gaussian curve to the $g-r$ color distribution of
  the resulting samples, redward of the distribution peak, and
  excluded all the galaxies that are bluer than the peak by more than
  1$\sigma$.
  Put differently, we selected an identical fraction of the reddest
  galaxies from each sample, assuming a normal distribution for both.
  However, since this color selection slightly changed the
  overall number density of the high redshift sample, we iterated over
  the last two steps until both criteria were satisfied, resulting in
  a matched galaxy number density of $4\pow10{-5}$ Mpc$^{-3}$.
  
  The initial selection criteria of the BOSS ``CMASS'' sources aim
  at finding galaxies at a roughly constant mass.
  This implies that the BOSS pipeline selects galaxies with a wider
  range of properties compared to the SDSS pipeline, including
  galaxies of late type morphology
  \citep[e.g.,][]{white_clustering_2011, masters_morphology_2011}.
  The color and luminosity that we apply on both BOSS and SDSS data
  sets minimize the contribution of such sources to our samples as the
  final selection only includes the reddest, most luminous galaxies at
  each redshift.
  We therefore refer to these galaxies as ``LRGs'' throughout the
  paper.
  We note that some contamination of dusty galaxies may still exist in
  the BOSS data set as simple color cuts likely do not completely
  exclude them \citep[e.g.,][]{bell_gems_2004, labbe_ultradeep_2003}.
    
  The color-magnitude distribution of the initial data sets can be
  seen in Figure \ref{fig:colmag} along with the selection
  cutoff values and resulting samples.
  The difference in cutoff restframe $M_g$ between the two redshift
  samples is less than 0.4 magnitudes and roughly corresponds to
  passive luminosity evolution.
  The final data set includes 12,813 galaxies at $0.6<z<0.7$ and 29,477 
  galaxies at $0.28<z<0.4$.

\section{Statistical derivation of the LRG satellite luminosity
  function}\label{sec:stder}
 \subsection{Region selection and object photometry\label{sec:regsel}}
  Each imaging field in SDSS is $811''\times 590''$ in angular size
  (see \citealp{stoughton_sloan_2002} for a description of the SDSS
  image processing), corresponding to a physical scale
  of roughly 5.6$\times$4.1 Mpc at $z=0.65$ and 3.9$\times$2.8 Mpc at
  $z=0.34$.
  We acquired $r-$ and $i-$band images from the survey archives for
  the selected galaxies in the low and high redshift samples,
  respectively, and extracted source photometry using SExtractor
  \citep{bertin_sextractor:_1996} in a 1000 kpc annulus around the
  LRGs ($145''$ at $z=0.65$ and $207''$ at $z=0.34$).
  We utilized a low detection threshold of 1$\sigma$ above the
  background, as determined by SExtractor, resulting in a large number
  of spurious detections of nise peaks in the images.
  However, since the random field catalogs contain the same false
  detections, this contribution to the luminosity function will be
  taken into account.
  We then calibrated the luminosity of each object using
  the K-correction value and luminosity distance of its adjacent LRG,
  assuming that all sources in any given field are at the same
  redshift.
  Although this assumption is grossly incorrect for any object that is
  not physically associated with the LRG itself, the large
  numbers of studied sources around the LRGs and in random fields
  ensure that each luminosity bin is well sampled in both data sets. 
  We discuss this further in Section \ref{sec:bgsub}.
  Finally, we divided the resulting all-inclusive catalogs into
  luminosity bins of size 0.01 dex, and summed the number of
  objects in each bin.
  A typical region selection at each redshift range can be seen in
  Figure \ref{fig:lf_example}.
  
 \subsection{Background and foreground subtraction\label{sec:bgsub}}
  Statistical subtraction of background and foreground sources in
  wide-field surveys can be performed locally or globally, with both
  methods providing successful and not inconsistent results
  (some discussion on the different approaches can be found in
  \citealp{goto_morphological_2003} and in \citealp{loh_color_2008}).
  We chose to use a local estimate of this contribution using source
  extraction and photometry in randomly-selected regions taken from
  the same SDSS fields in which the selected LRGs reside.
  This method better samples the large scale structure in which
  massive LRGs typically reside and therefore attempts to account for
  resulting line-of-sight overdensities.

%
  For each selected LRG we performed object photometry in a randomly
  positioned circular region, using identical thresholds as those used
  for the LRG fields (Section \ref{sec:regsel}).
  The large number of LRGs is well suited to this technique as it 
  ensures sufficient sampling of foreground and background objects.
  We allowed the random aperture centers to lie anywhere in the
  images such that the both LRG and random fields suffer from
  identical edge effects.
  This ensures that the cumulative size of all random
  apertures is the same as the total LRG field size.
  
  We then incorporated the measurements from all fields and binned
  them similarly to the LRG centered regions to derive a cumulative
  source light distribution of foreground and background sources.
  We repeated the last two steps until we had 10 random iterations for
  each LRG that did not overlap with the LRG aperture itself.
  We also allowed the random apertures to overlap with each other by
  no more than 30\% of the apertue area to ensure that the errors are
  not significantly correlated.
  
 \subsection{Photometric errors\label{sec:photerr}}
  Photometric measurements of the faintest detected objects may suffer
  from significant errors due to statistical variation in the
  background and other residual systematic biases (e.g., flat fielding
  issues, incorrect sky level determination).
  In order to estimate these errors we followed the technique
  suggested by \cite{labbe_ultradeep_2003} and utilized circular
  apertures to measure the total flux distribution in empty regions in
  both data sets.
  First we constructed a catalog of all the elliptical apertures
  that had been used by SExtractor to measure photometry from the
  images.
  We then calculated the average area of these apertures in each
  luminosity bin and produced a circular aperture of equivalent
  area.
  Next we masked out all the objects in each imaging field and
  measured the total flux in every aperture when placed at a random
  position in the field.
  We repeated the last step 1000 times and calculated the
  standard deviation of the resultant measurement distribution for
  each luminosity bin.
  
  \begin{figure*}
    \includegraphics[width=1.0\textwidth]{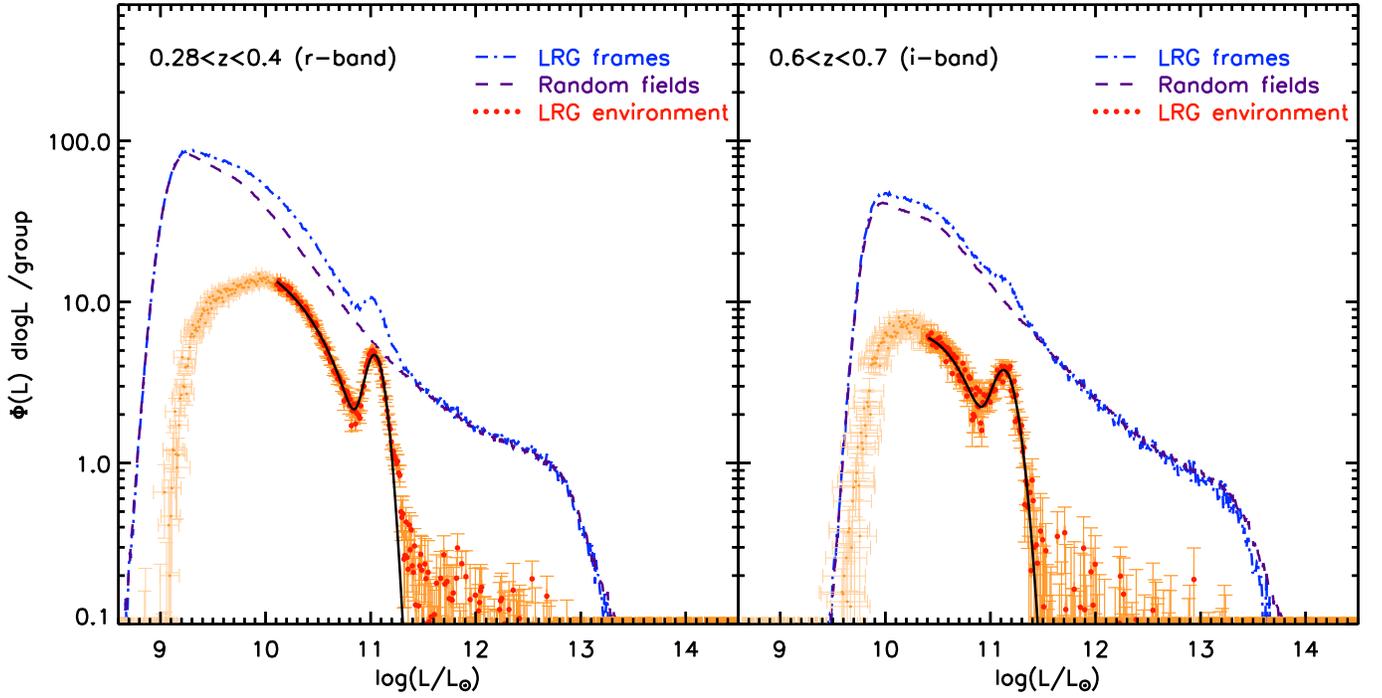}
    \caption{Galaxy luminosity functions of LRG satellites at average
      redshifts $z\sim0.34$ (left panel) and $z\sim0.65$ (right
      panel).
      The blue and purple curves represent the ``luminosity'' function
      of the LRG and random fields, respectively, and the red data
      points show the difference between the two.
      The black curve is a functional fit to the data at luminosities 
      brighter than the detection threshold while the pale orange points 
      show data at fainter luminosities.
      We note that the gap between the LRG and the most massive
      satellite galaxy is evident in both redshift bins.}
    \hfill
    \label{fig:lf_all}
  \end{figure*}

 \subsection{Luminosity functions}\label{sec:final}
  The galaxy luminosity functions that were derived in Section
  \ref{sec:regsel} are dominated by light from background and
  foreground sources in the LRG fields.
  In order to properly remove this contamination we averaged the
  random aperture measurements that were carried out in Section
  \ref{sec:bgsub}, in each luminosity bin, and subtracted them from
  the LRG measurements.
  We then normalized the resultant galaxy luminosity function by the
  total number of groups and clusters, assuming that every studied LRG
  is the central galaxy in its environment.
  A potential concern is that we may preferentially miss LRGs that are
  paired with other LRGs due to incompleteness of the SDSS redshift
  catalogs resulting from fiber collisions.
  We note that such pairs are rare, making up only a few percent of
  the sample at the selected luminosity range, after correcting for
  fiber collisions \citep[see e.g.][]{masjedi_very_2006,
  van_den_bosch_importance_2008, wake_2df-sdss_2008}.
  Therefore, the effect on the selection is small.
  Importantly, all LRGs that do not have a redshift measurement are
  included in our analysis as they are in the photometric catalogs
  that we used to construct the luminosity functions.


  In addition, we calculated the difference between SDSS \emph{model}
  and SExtractor \emph{AUTO} magnitudes and derived a correction
  factor for the latter.
  We applied this factor to the satellite functions assuming that SDSS
  model magnitudes better assess the total flux from satellite
  galaxies.
  We further discuss this in the appendix.
  Finally, we applied the same filter transformations discussed in
  Section \ref{ndens} to the luminosity functions and normalized the
  resultant measurements to solar luminosities using
  $M_{\odot,g}=5.12$ \citep[in the AB 
  system,][]{blanton_<i>k</i>_2007}.

  The calibrated LRG satellite luminosity functions are shown in
  Figure \ref{fig:lf_all}.
  The blue dot-dashed and the purple dashed lines show the observed
  luminosity functions of the LRG and random fields, respectively.
  The red data points show the difference between the two curves which
  is henceforth assumed to represent the luminosity function of
  galaxies associated with LRG environments.
  These luminosity functions also include the central LRGs themselves,
  for which photometry was extracted in an identical way to all other
  sources in the field.
  The vertical error bars represent 1$\sigma$ of the distribution of
  random measurements in each luminosity bin (Section
  \ref{sec:bgsub}) and the horizontal error bars show the standard
  deviation of 1000 empty aperture measurements (Section
  \ref{sec:photerr}).

  The luminosity functions are markedly different from a Schechter
  function and show a clear peak at the luminosity of the LRG at
  both redshifts.
  This peak is even visible in the observed luminosity function (the
  blue dot-dashed curve which is not corrected for foreground and
  background objects).
  The fact that a large fraction of the luminosity in a typical LRG
  environment is locked in the central galaxy itself is the key result
  of this study.
  In the following sections we will quantify the luminosity functions
  and the total luminosity in the LRG and in its satellite galaxies.

\section{Quantifying the luminosity function of LRG
  environments}\label{sec:quant}
 \subsection{Functional fits}\label{sec:ffits}
  The derived luminosity functions that are presented in Figure
  \ref{fig:lf_all} are very different from a Schechter distribution,
  in agreement with halo-based studies of nearby and distant
  environments \citep[e.g.,][]{yang_galaxy_2008, brown_red_2008}.
  Following \cite{yang_galaxy_2008} we instead quantified the galaxy
  luminosity distributions using a two-component fit:
  \begin{equation}
    \Phi(L)=\Phi_{LRG}(L)+\Phi_{sat}(L),
  \end{equation}
  where $\Phi_{LRG}$ and $\Phi_{sat}$ are the distributions of
  LRG and satellite luminosities.
  For the LRG luminosities we assumed a log-normal distribution,
  \begin{equation}
    \Phi_{LRG}(L) = \frac{A}{\sqrt{2\pi}\sigma_c}
    \exp\left[\frac{-\left(\log L-\log L_c\right)^2}{2\sigma_c^2}\right],
  \end{equation}
  where the number of LRGs per group or cluster $A$ is set to 1 and
  the distribution width $\sigma_c$ and peak center $\log L_c$ are
  left as free parameters.
  For the satellite luminosity distribution we used a
  \cite{schechter_analytic_1976} function:
  \begin{equation}
    \Phi_{sat}(L) = \phi_s \left(\frac{L}{L_s}\right)^{(\alpha_s+1)}
    \exp \left[-\frac{L}{L_s}\right].
  \end{equation}
  Attempts at fitting the satellite luminosity distribution with a
  modified Schechter function, as suggested by
  \cite{yang_galaxy_2008}, resulted in worse agreement with the data. 
  We fit all five free parameters ($\sigma_c$, $\log L_c$, $\phi_s$,
  $\log L_s$ and $\alpha_s$) simultaneously using the non-linear least
  squares curve fitting program MPFIT
  \citep{markwardt_non-linear_2009}.
  Since our data quickly become incomplete below some flux threshold
  we restricted the functional fits to luminosities higher than the
  turnover luminosity of $\log L_{min} = 10.1$ and $10.4$ in the low
  and high redshift bins.
  The best-fit parameters are given in Table \ref{tab:fits} and the
  resulting luminosity function curves are shown in Figure
  \ref{fig:lf_all} as black solid lines.
  The slope of the power-law part of the Schechter
  function, represented by the parameter $\alpha_s$, is not well
  constrained and is sensitive to the adopted value of $L_{min}$.
  When varying $\log(L_{min})$ from 10.1 to 10.6 for the low-redshift
  sample the slope changes by 0.4 dex.
  To better estimate the faint end of the Schechter function we
  repeated the same model fitting for LRGs in Stripe 82.

  \begin{table}
    \caption{Best-fit parameters of the galaxy luminosity function.}
    \centering
    \begin{tabular}{l c c c}
      \hline\hline
       & SDSS & BOSS & Stripe 82\\
       & $0.28<z<0.40$ & $0.60<z<0.70$ & $0.28<z<0.40$\\
      \hline
      $A$ (fixed) & 1.00 & 1.00 & 1.00 \\
      $\sigma_c$ & 0.09 & 0.12 & 0.09 (fixed)\\
      $\log L_c$ & 11.04 & 11.14 & 11.00\\
      $\phi_s$ & 18.6 & 15.3 & 24.3\\ 
      $\log L_s$ & 10.5 & 10.5 & 10.4\\
      $\alpha_s$ & -1.11$\pm$0.47\footnote{Errors were calculated
       by varying the analysis threshold $\log(L_{min})$ by 0.5 dex.}
       & -0.46$\pm$1.15
       & -0.95$\pm$0.18\\
      \hline\hline
      \label{tab:fits}      
    \end{tabular}
  \end{table}
    
\begin{figure}
    \begin{center}
      \includegraphics[width=0.47\textwidth]{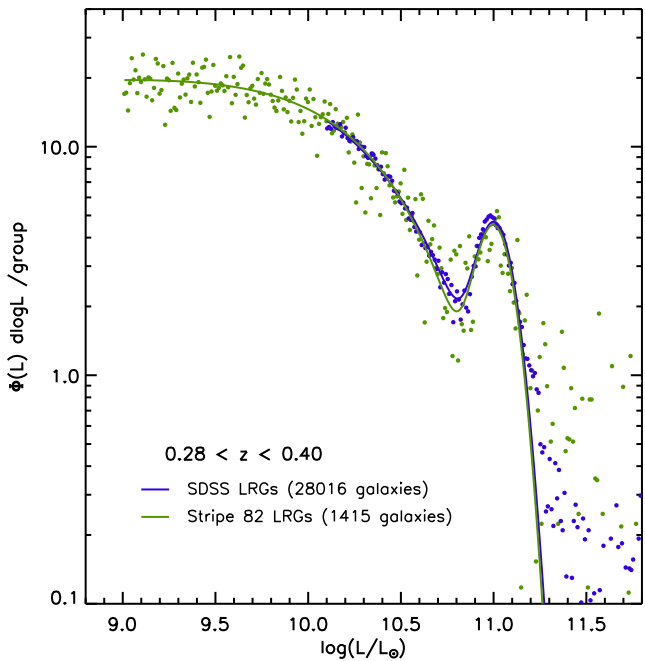}
      \caption{Comparison between the luminosity functions derived from
	individual SDSS LRG frames (blue data points) and from deep
	Stripe 82 stacks (green data points) in the redshift range
	$0.28<z<0.40$.
	Solid lines are functional fits to the data using the
	two-component model described in Section \ref{sec:stder}.
	The faint-end slope of the Schechter function can be reliably
	measured and it has a value of $-0.95$.
      }
      \label{fig:s82}
    \end{center}
  \end{figure}

  Stripe 82 is a narrow region in the sky that was imaged multiple
  times as part of SDSS under a variety of observing conditions
  \cite[e.g.,][]{abazajian_seventh_2009}.
  Individual images of the same fields were summed to create deep
  co-added frames of this region, each made of $11$ to $75$ exposures
  (with a median of $52$ exposures).
  We repeated the analysis described in Section \ref{sec:stder} in
  fields containing all 1415 Stripe 82 LRGs at $0.28<z<0.40$
  (redshifts were measured as part of the main SDSS survey).
  For the functional fits we kept the parameter $\sigma_c$ fixed
  since the relatively small number of LRGs in Stripe 82 implies that
  the gap region of the luminosity function is not well constrained.
  To calibrate the photometry we matched the resulting luminosity of
  the LRGs themselves to that of the same LRGs in the SDSS frames.
  Figure \ref{fig:s82} shows a comparison between the luminosity
  function that was derived from individual SDSS frames (blue points)
  and the one derived from the deep Stripe 82 stacks (green points).
  As expected, there is good agreement between the two curves down to
  the SDSS threshold luminosity of $\log{L}\sim10.1 L\solar$.
  The faint-end slope of the Schechter function is better constrained 
  from the deep Stripe 82 data and it has a value of $-0.95$.

 \subsection{Measurements of the gap width}\label{meas_gw}
  The most outstanding feature in both galaxy luminosity functions
  that are presented in Figure \ref{fig:lf_all} is a gap at the bright
  end between the LRG luminosity and that of the most luminous
  satellites. 
  Similar luminosity gaps in nearby ($z<0.2$) massive groups and
  clusters are typically interpreted as a proxy of the magnitude
  difference between the first and second most luminous group members
  \citep[e.g.,][]{ponman_possible_1994, yang_galaxy_2008}.
  However, the statistical nature of this study and the methods used
  to derive the galaxy luminosity functions imply that we cannot
  measure such a magnitude difference for any given group or cluster
  as membership is not assigned to individual sources.
  Instead, we treat the gap in the luminosity function as a
  probability distribution for finding satellites at a given relative
  luminosity compared to the central.
  Thus, we quantify the magnitude gap by finding the luminosity
  above which LRG groups and clusters have on average exactly one
  satellite.
  The ratio between this luminosity and the peak of LRG luminosities
  is then roughly equivalent to the magnitude gap between the two most 
  luminous members of the environment.
  Put differently, we calculate the luminosity $L_u$ above which the
  integrand over the satellite luminosity distribution equals unity:
  \begin{equation}
    \int_{L_u}^{\infty}\phi_s
    \left(\frac{L}{L_s}\right)^{(\alpha_s+1)} \exp
    \left[-\frac{L}{L_s}\right]d\log{L}=1
  \end{equation}
  This statistic implies a gap width of $\log{L_c}-\log{L_u} \sim
  0.5$ dex, or roughly 1.3 magnitudes, at both redshifts.
  An alternative measure of the gap width is the difference between
  $\log{L_c}$ and $\log{L_s}$ which is also consistent between the two
  redshift samples and has a value of roughly 0.5 dex.
  We note, however, that this measurement is less robust as the
  parameter $\log{L_s}$ is degenerate with the other parameters of the
  Schechter function.
  
\section{Discussion}
 \subsection{Evidence for an early formation of LRG environments}
  Measurements of the magnitude difference between the first and
  second most luminous group members have typically been interpreted
  as a gauge of the group age
  \citep[e.g.,][]{sandage_redshift-distance_1973, tremaine_test_1977,
    schneider_ccd_1983, barnes_evolution_1989, ponman_possible_1994,
    khosroshahi_old_2004, donghia_formation_2005,
    milosavljevi_cluster_2006, van_den_bosch_towards_2007,
    yang_galaxy_2008, dariush_mass_2010}.
  In this model the most massive members merge quickly (within a
  few tenths of a Hubble time) and leave behind only significantly
  less massive satellites.
  \cite{masjedi_growth_2008} showed supporting evidence for this model
  by deriving the small scale correlation function of SDSS LRGs and
  estimating that these central galaxies can only grow by up to a few
  percent via merging with their satellites.  
  The galaxy luminosity functions presented in figure \ref{fig:lf_all}
  further support this and suggest that LRGs typically live in groups
  where the central galaxy is significantly more massive than its most
  luminous satellite.
  
\begin{figure}
   \includegraphics[width=0.47\textwidth]{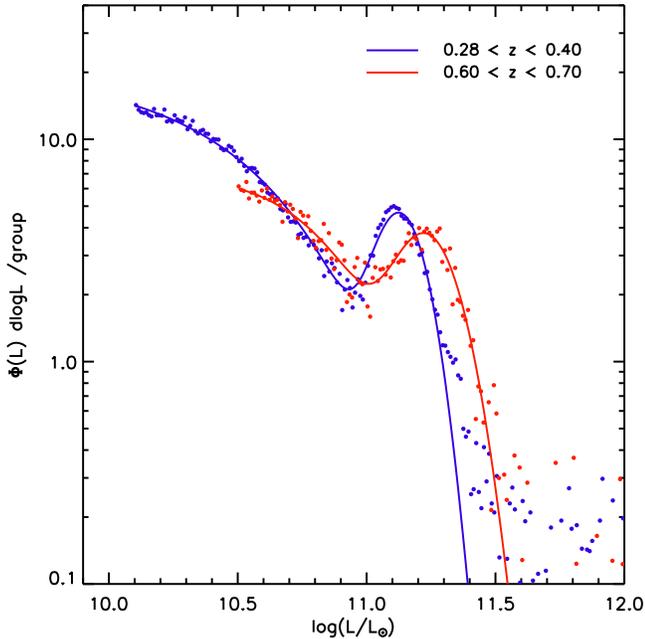}
   \caption{Luminosity distributions of the two redshift bins
     over plotted with their respective best fit models.
     The gap in the luminosity function is evident, and generally
     similar in properties, at both redshifts.
     The ratio of LRG peak luminosities is consistent with the
     luminosity evolution model of \cite{blanton_galaxy_2003}.}
   \hfill
   \label{fig:lfmodels}
 \end{figure}

  In addition, Figure \ref{fig:lfmodels} shows that the depth of the
  luminosity gap at $z\sim0.34$, as well as its width, are generally
  similar to the gap properties at $z\sim0.65$.
  This is consistent with no significant evolution in the gap
  properties between the two redshift bins and therefore no
  significant merger activity between six and four billion years ago.
  Moreover, the existence of the gap at $z=0.67$ is consistent with
  the scenario in which the gap (and group) formed early.

  We note that the ratio between the two LRG peaks is consistent with
  passive luminosity evolution as described by the model derived by
  \cite{blanton_galaxy_2003} for the same galaxies in SDSS and is in
  agreement with \cite{tojeiro_evolution_2010}.
  
 \subsection{Mergers and the mild mass evolution of LRGs}
\begin{figure}
    \includegraphics[width=0.47\textwidth]{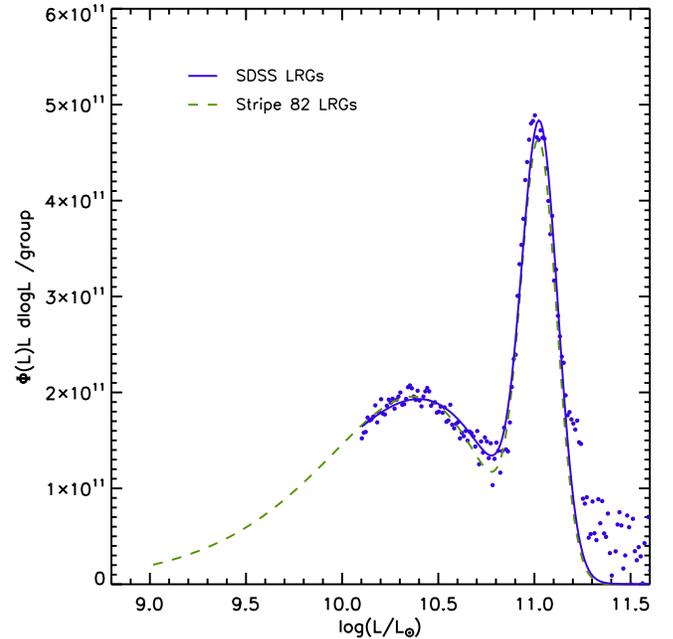}
    \caption{Total luminosity per log luminosity bin for
      the low-redshift sample.
      This figure shows that the ratio between the peaks of LRG and
      satellite galaxy luminosity distributions is roughly 4:1.
      This ratio implies that most of the satellite mass is in
      galaxies that are roughly four times less massive than the LRG
      itself.
    }
    \hfill
    \label{fig:lphi}
  \end{figure}

  The luminosity functions derived in Sections \ref{sec:stder} and
  \ref{sec:quant} can be used to directly characterize the reservoir
  of satellite galaxies with which the central LRG can merge.
  We utilize these luminosity distributions to calculate the mass of
  satellite galaxies which contribute the most to the LRG stellar mass
  growth.
  In Figure \ref{fig:lphi} we present the distribution of total galaxy
  luminosities per log luminosity bin for the low-redshift sample.
  From this relation we can measure the ratio between the peak of LRG
  and satellite galaxy luminosity distributions.
  This ratio has a value 4:1 and it suggests that the satellite mass
  distribution peaks at 25\% of the LRG mass.
  This is consistent with the derivations of gap width (Section
  \ref{meas_gw}) which imply that the most luminous LRG satellite is
  on average more than three times fainter than the LRG itself.
  This result supports a scenario in which major mergers within LRG
  environments are improbable and that any mass growth takes
  place through mergers with mass ratios of 1:4 or lower
  \citep[e.g.,][]{masjedi_growth_2008, kaviraj_role_2009}.
  
  In addition, by separately integrating the total LRG and satellite
  luminosity distributions of Figure \ref{fig:lphi} we calculate that
  roughly 35\% of the mass in LRG environments is locked in the
  central galaxy itself.
  Although the potential for stellar mass growth via mergers within
  groups or clusters is not insignificant, significant evolution, when
  restricted to growth through minor mergers, may in fact require many
  Gyrs.
  For comparison, in massive clusters, where the galaxy luminosity
  distribution is better described by a Schechter function, the
  central galaxy contains as little as 15\% or less of the total
  stellar mass in the environment.
  In such environments the central galaxy can grow quickly in mass
  through major mergers with its most massive satellites.

  We note that this estimate is an upper limit for the maximum mass
  growth through mergers within the LRG group.
  Since the colors of satellite galaxies are expected to be bluer than
  the those of central LRGs, a more accurate mass calibration will
  likely shift the satellite mass function to lower masses and
  increase the luminosity gap width.
  Proper estimates of the color of satellites in LRG group are
  therefore important for this analysis but they are unfortunately
  beyond the scope of this study.
  
 \subsection{Uncertainties in mass estimates}\label{sec:uncert}
  We note that the luminosity, and therefore mass, calibrations of
  the luminosity functions that are presented in Figures
  \ref{fig:lf_all} and \ref{fig:lfmodels} are only as valid as some of
  the assumptions that were made for the functional fits.
  For example, in order to calibrate the luminosity functions we
  assumed that LRG satellites typically have the same colors as the
  central galaxy.
  If the satellites are instead bluer than the LRG, their calibrated
  luminosities would be fainter, implying lower total stellar mass.
  In that case the mass function of LRG environments would exhibit an
  even wider gap between the two most massive members.
  Although group satellites have indeed been observed to be bluer
  than a typical LRG, most of them are close in color to their
  central galaxy \citep[e.g.,][]{yang_galaxy_2008,
  balogh_colour_2009, weinmann_environmental_2009}.

  An additional error may arise from the magnitude correction that was
  derived in Section \ref{sec:final} and is presented in Figure
  \ref{fig:s82}.
  This estimate relies on the assumption that the magnitude difference
  between the two photometric measurements remains constant even at
  fainter brightnesses.
  An increase in this correction factor at fainter magnitudes may
  imply that we underestimate the luminosity of satellite galaxies and
  therefore potentially overestimate the width of the luminosity gap.

  Finally, it is possible that sources that are at a small angular
  distance from the LRGs are missed from our catalogs due to
  imperfect source separation by SExtractor.
  Although this could also potentially lead to an overestimate of the
  gap width it is more likely that sources that are missed are faint
  and do not contribute much to the luminosity functions in the gap
  region. 
  
  \begin{figure}
    \includegraphics[width=0.47\textwidth]{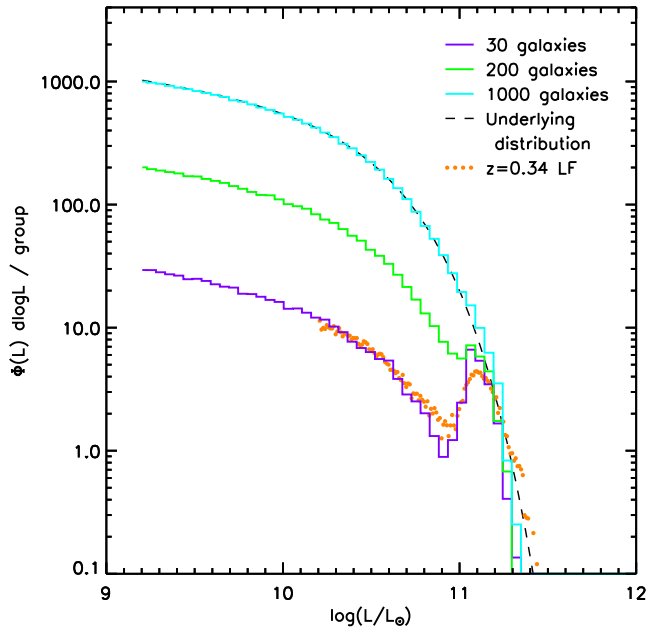}
    \caption{Simulated galaxy luminosity functions for group and
      cluster sizes of 30,200 and 1000 galaxies per environment.
      Galaxy luminosities were drawn at random from an assumed
      underlying distribution which is modeled as a Schechter function.
      Each curve represents the average of 1000 groups and clusters
      where at least one galaxy is more luminous than a threshold of
      $logL=11$.
      The shape of the observed galaxy luminosity function is highly
      correlated with the number of satellites even for a constant
      underlying distribution, resulting from the steepness of this
      distribution at high luminosities.
      Also plotted is the derived luminosity function for the SDSS
      data set of LRGs at $0.28<z<0.40$.}
    \hfill
    \label{fig:plot_gap}
  \end{figure}
  
 \subsection{The Schechter function and luminosity gap}
  Gaps in the luminosity functions of groups and clusters are often
  observed in studies of the environments of luminous galaxies
  \citep[e.g.,][]{yang_galaxy_2008, brown_red_2008}.
  These gaps may be created during the initial formation of 
  massive centrals by mergers of the most massive members in an
  environment.
  Such (major) mergers can increase the luminosity of the central
  galaxy while at the same time effectively decrease the luminosity
  of the most massive satellite.
  For example, a sparse group hosting two $L^{\star}$ galaxies would
  not be selected for this study and it would also not have a gap in
  its luminosity function.
  However, some time after the two galaxies merge the remnant galaxy
  will likely be luminous enough to be classified as an LRG and a
  gap in the luminosity function will appear.  
  In this scenario, luminous massive galaxies in sparse
  groups are  not likely to have any remaining massive companions in
  their environments.
  It therefore follows that group LRGs preferentially live in
  environments where a luminosity gap can be observed even if the
  underlying luminosity distribution does not exhibit a gap.
  
  Here we test whether the observed gap can be a result of the
  selection of the LRGs, combined with sparse sampling of an
  underlying Schechter function.
  We start by assuming that this distribution is the same in all
  environments, regardless of other properties, and that it can be
  well described by a Schechter function.
  We then randomly sample the assumed Schechter distribution $N$
  times, where each reading represents a galaxy observation and $N$ is
  the total number of galaxies in a ``group''.
  We record the resultant galaxy luminosities and derive luminosity
  functions for all groups.
  Finally, we repeat the last two steps until at least 1000 groups (or
  clusters) of each size $N$ have one or more luminous galaxies with
  $\log(L)\ge11 L\solar$ and exclude all other groups.
  Figure \ref{fig:plot_gap} shows the average luminosity function of
  the 1000 random samples for group sizes of $N=30$, 200 and 1000
  galaxies.
  The underlying distribution from which galaxies were sampled
  is also plotted (black dashed line).
  The luminosity gap can be seen in the average distribution of groups
  with 30 and 200 galaxies, but not in the richest clusters.
  
  It is therefore evident that the gap in the luminosity function may
  indeed result from the criteria used to select our sample.
  The steep bright end of the Schechter function implies that the
  probability of having multiple luminous galaxies in one environment
  decreases with the number of physically associated members.
  It then follows that LRG environments are not necessarily unique in
  their underlying luminosity distribution but rather in that they
  host a luminous galaxy.
  Furthermore, this may suggest that groups that do not host a bright
  galaxy can easily be missed by group finding studies as all of the
  member galaxies may be undetected.
  We note that this result is not restricted to a Schechter
  distribution and that any underlying distribution which is much
  steeper at its bright end may reproduce a similar luminosity gap.

  Consequently, if environments are selected by their total halo mass
  rather than the luminosity of their most massive member, they may
  exhibit a much shallower gap or even none at all.
  When assuming a universal underlying Schechter distribution for all
  halos of a given mass, groups hosting a massive galaxy become rare
  and their contribution to the average luminosity function
  diminishes.
  In fact, in our simulations such groups account for only 3\% of the
  total environments with the same number of satellites.
  We note, however, that these simulations are over-simplistic and that 
  they do not represent the true global distribution of halo masses.
  Moreover, all groups hosting the same number of galaxies do not have
  the same total mass, suggesting that this value may be a gross 
  underestimate of the fraction of groups hosting an LRG.
  Nevertheless, any selection which is based on the central galaxy 
  luminosity would likely underestimate the scatter in the masses
  of centrals in similar halos.
  Such a scatter was observed by \cite{yang_galaxy_2008}, who noticed
  that the distribution of central galaxy luminosities $(\sigma_c)$
  has a constant width of roughly 0.15 dex for
  $M_h>10^{13}h^{-1}M\solar$.
  Moreover, some scatter is required to provide good halo model fits
  to the clustering and space density of LRGs
  \citep{wake_2df-sdss_2008, zheng_halo_2009, white_clustering_2011}.
  
\section{Summary and conclusions}
 In this paper we utilized imaging data of more than 40,000 SDSS LRGs
 in the redshift ranges $0.28<z<0.40$ and $0.60<z<0.70$ to study the
 luminosity function of LRG satellite galaxies.
 We extracted source photometry in 1000 kpc apertures around the LRGs
 themselves, as well as in randomly selected fields, to characterize
 and remove background and foreground contamination in a
 statistical way.
 The large size of SDSS makes it an ideal data set for this study as
 it reduces statistical noise and allows for excellent determination
 of the variation between fields.
 The addition of BOSS spectra to the imaging data allowed us to study
 the galaxy luminosity function around LRGs out to $z=0.7$ for the
 first time.
 
 We successfully determined the galaxy luminosity function of LRG
 satellites down to two magnitudes fainter than the LRG brightness at
 $z\sim0.65$ despite the relatively shallow depth of SDSS.
 We did so without acquiring additional spectra to confirm member
 candidates and by assuming that the properties of LRG satellites
 are correlated with the properties of the LRGs themselves.
 This technique can be utilized to study the environment around any
 type of galaxy in a well-defined sample and in a sufficiently large
 data set. 

 We found that the luminosity function of LRG environments is markedly
 different from a Schechter function and that LRGs are typically $\sim
 1.3$ magnitudes brighter than the next brightest galaxy in their
 environment.
 We showed that most of the group mass that is not in the central
 itself is concentrated in satellite galaxies that are roughly four
 times less luminous than the LRG.
 This implies that major mergers within LRG groups and clusters are
 improbable and that any stellar mass growth likely occurs through
 minor mergers. 
 We also demonstrated that the luminosity function gap is already
 in place at $z=0.65$ and estimated that the total mass in LRGs
 accounts for roughly 35\% of the total stellar mass in their
 environment.
 The existence of the gap at this redshift supports that
 LRG environments typically formed early and were already in place 6
 Gyrs in the past.

 Lastly, we performed simple assembly simulations of galaxy groups and
 clusters and demonstrated that the selection criteria used in this
 study may have preferentially picked environments that have a gap in
 their luminosity function.
 In this case the luminosity gap results from the properties of the
 underlying Schechter function that was used in the simulations and
 from the requirement to have a luminous galaxy in every selected
 environment. 
 This further implies that the LRG luminosity function may not be
 inherently unique and that a true halo mass selected sample may
 exhibit a large scatter in the properties of its most massive
 galaxies. 
 
\begin{acknowledgements}
  We are grateful to Daniel Eisenstein for an engaging discussion
  which helped improve the manuscript.
  We also thank the referee for a very constructive report, which
  improved the paper.  
  Support from the CT Space Grant is gratefully acknowledged.
  
  Funding for SDSS-III has been provided by the Alfred P. Sloan
  Foundation, the Participating Institutions, the National Science
  Foundation, and the U.S. Department of Energy. The SDSS-III web site
  is http://www.sdss3.org/.

  SDSS-III is managed by the Astrophysical Research Consortium for the
  Participating Institutions of the SDSS-III Collaboration including
  the University of Arizona, the Brazilian Participation Group,
  Brookhaven National Laboratory, University of Cambridge, University
  of Florida, the French Participation Group, the German Participation
  Group, the Instituto de Astrofisica de Canarias, the Michigan
  State/Notre Dame/JINA Participation Group, Johns Hopkins University,
  Lawrence Berkeley National Laboratory, Max Planck Institute for
  Astrophysics, New Mexico State University, New York University, Ohio
  State University, Pennsylvania State University, University of
  Portsmouth, Princeton University, the Spanish Participation Group,
  University of Tokyo, University of Utah, Vanderbilt University,
  University of Virginia, University of Washington, and Yale
  University.  
\end{acknowledgements}
 
\bibliography{ms}

\begin{thebibliography}{103}
\expandafter\ifx\csname natexlab\endcsname\relax\def\natexlab#1{#1}\fi

\bibitem[{Abazajian {et~al.}(2009)Abazajian, Allende~Prieto, An, Anderson,
  Anderson, Annis, Bahcall, {Bailer-Jones}, Barentine, Bassett, Becker, Beers,
  Bell, Belokurov, Berlind, Berman, Bernardi, Bickerton, Bizyaev, Blakeslee,
  Blanton, Bochanski, Boroski, Brewington, Brinchmann, Brinkmann, Brunner,
  Budavári, Carey, Carliles, Carr, Castander, Cinabro, Connolly, Csabai,
  Cunha, Czarapata, Davenport, de~Haas, Dilday, Doi, Eisenstein, Evans, Evans,
  Fan, Friedman, Frieman, Fukugita, Gänsicke, Gates, Gillespie, Gilmore,
  Gonzalez, Gonzalez, Grebel, Gunn, Györy, Hall, Harding, Harris, Harvanek,
  Hawley, Hayes, Heckman, Hendry, Hennessy, Hindsley, Hoblitt, Hogan, Hogg,
  Holtzman, Hyde, Ichikawa, Ichikawa, Im, Ivezić, Jester, Jiang, Johnson,
  Jorgensen, Jurić, Kent, Kessler, Kleinman, Knapp, Konishi, Kron, Krzesinski,
  Kuropatkin, Lampeitl, Lebedeva, Lee, Lee, Leger, Lépine, Li, Lima, Lin,
  Long, Loomis, Loveday, Lupton, Magnier, Malanushenko, Malanushenko,
  Mandelbaum, Margon, Marriner, {Martínez-Delgado}, Matsubara, {McGehee},
  {McKay}, Meiksin, Morrison, Mullally, Munn, Murphy, Nash, Nebot, Neilsen,
  Newberg, Newman, Nichol, Nicinski, {Nieto-Santisteban}, Nitta, Okamura,
  Oravetz, Ostriker, Owen, Padmanabhan, Pan, Park, Pauls, Peoples, Percival,
  Pier, Pope, Pourbaix, Price, Purger, Quinn, Raddick, Fiorentin, Richards,
  Richmond, Riess, Rix, Rockosi, Sako, Schlegel, Schneider, Scholz, Schreiber,
  Schwope, Seljak, Sesar, Sheldon, Shimasaku, Sibley, Simmons, Sivarani, Smith,
  Smith, Smolčić, Snedden, Stebbins, Steinmetz, Stoughton, Strauss,
  Subba~Rao, Suto, Szalay, Szapudi, Szkody, Tanaka, Tegmark, Teodoro, Thakar,
  Tremonti, Tucker, Uomoto, Vanden~Berk, Vandenberg, Vidrih, Vogeley, Voges,
  Vogt, Wadadekar, Watters, Weinberg, West, White, Wilhite, Wonders, Yanny,
  Yocum, York, Zehavi, Zibetti, \& Zucker}]{abazajian_seventh_2009}
Abazajian, K.~N., {et~al.} 2009, The Astrophysical Journal Supplement Series,
  182, 543

\bibitem[{Balogh {et~al.}(2004)Balogh, Baldry, Nichol, Miller, Bower, \&
  Glazebrook}]{balogh_bimodal_2004}
Balogh, M.~L., Baldry, I.~K., Nichol, R., Miller, C., Bower, R., \& Glazebrook,
  K. 2004, The Astrophysical Journal, 615, L101

\bibitem[{Balogh {et~al.}(2009)Balogh, {McGee}, Wilman, Bower, Hau, Morris,
  Mulchaey, Oemler, Parker, \& Gwyn}]{balogh_colour_2009}
Balogh, M.~L., {et~al.} 2009, Monthly Notices of the Royal Astronomical
  Society, 398, 754

\bibitem[{Barnes(1989)}]{barnes_evolution_1989}
Barnes, J.~E. 1989, Nature, 338, 123

\bibitem[{Bell {et~al.}(2006)Bell, Phleps, Somerville, Wolf, Borch, \&
  Meisenheimer}]{bell_merger_2006}
Bell, E.~F., Phleps, S., Somerville, R.~S., Wolf, C., Borch, A., \&
  Meisenheimer, K. 2006, The Astrophysical Journal, 652, 270

\bibitem[{Bell {et~al.}(2004)Bell, {McIntosh}, Barden, Wolf, Caldwell, Rix,
  Beckwith, Borch, Häussler, Jahnke, Jogee, Meisenheimer, Peng, Sanchez,
  Somerville, \& Wisotzki}]{bell_gems_2004}
Bell, E.~F., {et~al.} 2004, The Astrophysical Journal Letters, 600, L11

\bibitem[{Bertin \& Arnouts(1996)}]{bertin_sextractor:_1996}
Bertin, E., \& Arnouts, S. 1996, Astronomy and Astrophysics Supplement Series,
  117, 393

\bibitem[{Bezanson {et~al.}(2009)Bezanson, van Dokkum, Tal, Marchesini, Kriek,
  Franx, \& Coppi}]{bezanson_relation_2009}
Bezanson, R., van Dokkum, P.~G., Tal, T., Marchesini, D., Kriek, M., Franx, M.,
  \& Coppi, P. 2009, Astrophysical Journal, 697, 1290

\bibitem[{Blanton \& Roweis(2007)}]{blanton_<i>k</i>_2007}
Blanton, M.~R., \& Roweis, S. 2007, The Astronomical Journal, 133, 734

\bibitem[{Blanton {et~al.}(2003)Blanton, Hogg, Bahcall, Brinkmann, Britton,
  Connolly, Csabai, Fukugita, Loveday, Meiksin, Munn, Nichol, Okamura, Quinn,
  Schneider, Shimasaku, Strauss, Tegmark, Vogeley, \&
  Weinberg}]{blanton_galaxy_2003}
Blanton, M.~R., {et~al.} 2003, The Astrophysical Journal, 592, 819

\bibitem[{Bolzonella {et~al.}(2010)Bolzonella, Kovač, Pozzetti, Zucca,
  Cucciati, Lilly, Peng, Iovino, Zamorani, Vergani, Tasca, Lamareille, Oesch,
  Caputi, Kampczyk, Bardelli, Maier, Abbas, Knobel, Scodeggio, Carollo,
  Contini, Kneib, Le~Fèvre, Mainieri, Renzini, Bongiorno, Coppa, de~la Torre,
  de~Ravel, Franzetti, Garilli, Le~Borgne, Le~Brun, Mignoli, Pelló,
  {Perez-Montero}, Ricciardelli, Silverman, Tanaka, Tresse, Bottini, Cappi,
  Cassata, Cimatti, Guzzo, Koekemoer, Leauthaud, Maccagni, Marinoni,
  {McCracken}, Memeo, Meneux, Porciani, Scaramella, Aussel, Capak, Halliday,
  Ilbert, Kartaltepe, Salvato, Sanders, Scarlata, Scoville, Taniguchi, \&
  Thompson}]{bolzonella_tracking_2010}
Bolzonella, M., {et~al.} 2010, Astronomy and Astrophysics, 524, 76

\bibitem[{Bournaud {et~al.}(2007)Bournaud, Jog, \&
  Combes}]{bournaud_multiple_2007}
Bournaud, F., Jog, C.~J., \& Combes, F. 2007, Astronomy and Astrophysics, 476,
  1179

\bibitem[{{Boylan-Kolchin} {et~al.}(2006){Boylan-Kolchin}, Ma, \&
  Quataert}]{boylan-kolchin_red_2006}
{Boylan-Kolchin}, M., Ma, C., \& Quataert, E. 2006, Monthly Notices of the
  Royal Astronomical Society, 369, 1081

\bibitem[{Brown {et~al.}(2008)Brown, Zheng, White, Dey, Jannuzi, Benson, Brand,
  Brodwin, \& Croton}]{brown_red_2008}
Brown, M. J.~I., {et~al.} 2008, The Astrophysical Journal, 682, 937

\bibitem[{Bruzual \& Charlot(2003)}]{bruzual_stellar_2003}
Bruzual, G., \& Charlot, S. 2003, Monthly Notices of the Royal Astronomical
  Society, 344, 1000

\bibitem[{Bundy {et~al.}(2009)Bundy, Fukugita, Ellis, Targett, Belli, \&
  Kodama}]{bundy_greater_2009}
Bundy, K., Fukugita, M., Ellis, R.~S., Targett, T.~A., Belli, S., \& Kodama, T.
  2009, The Astrophysical Journal, 697, 1369

\bibitem[{Bundy {et~al.}(2006)Bundy, Ellis, Conselice, Taylor, Cooper, Willmer,
  Weiner, Coil, Noeske, \& Eisenhardt}]{bundy_mass_2006}
Bundy, K., {et~al.} 2006, The Astrophysical Journal, 651, 120

\bibitem[{Chiboucas {et~al.}(2010)Chiboucas, Tully, Marzke, Trentham, Ferguson,
  Hammer, Carter, \& Khosroshahi}]{chiboucas_keck/lris_2010}
Chiboucas, K., Tully, R.~B., Marzke, R.~O., Trentham, N., Ferguson, H.~C.,
  Hammer, D., Carter, D., \& Khosroshahi, H. 2010, The Astrophysical Journal,
  723, 251

\bibitem[{Christlein \& Zabludoff(2003)}]{christlein_galaxy_2003}
Christlein, D., \& Zabludoff, A.~I. 2003, The Astrophysical Journal, 591, 764

\bibitem[{Coil {et~al.}(2006)Coil, Gerke, Newman, Ma, Yan, Cooper, Davis,
  Faber, Guhathakurta, \& Koo}]{coil_deep2_2006}
Coil, A.~L., {et~al.} 2006, The Astrophysical Journal, 638, 668

\bibitem[{Dariush {et~al.}(2010)Dariush, Raychaudhury, Ponman, Khosroshahi,
  Benson, Bower, \& Pearce}]{dariush_mass_2010}
Dariush, A.~A., Raychaudhury, S., Ponman, T.~J., Khosroshahi, H.~G., Benson,
  A.~J., Bower, R.~G., \& Pearce, F. 2010, Monthly Notices of the Royal
  Astronomical Society, 405, 1873

\bibitem[{{D'Onghia} {et~al.}(2005){D'Onghia}, {Sommer-Larsen}, Romeo, Burkert,
  Pedersen, Portinari, \& Rasmussen}]{donghia_formation_2005}
{D'Onghia}, E., {Sommer-Larsen}, J., Romeo, A.~D., Burkert, A., Pedersen, K.,
  Portinari, L., \& Rasmussen, J. 2005, The Astrophysical Journal, 630, L109

\bibitem[{Driver {et~al.}(1994)Driver, Phillipps, Davies, Morgan, \&
  Disney}]{driver_dwarf_1994}
Driver, S.~P., Phillipps, S., Davies, J.~I., Morgan, I., \& Disney, M.~J. 1994,
  Monthly Notices of the Royal Astronomical Society, 268, 393

\bibitem[{Eisenstein {et~al.}(2001)Eisenstein, Annis, Gunn, Szalay, Connolly,
  Nichol, Bahcall, Bernardi, Burles, Castander, Fukugita, Hogg, Ivezić, Knapp,
  Lupton, Narayanan, Postman, Reichart, Richmond, Schneider, Schlegel, Strauss,
  {SubbaRao}, Tucker, Vanden~Berk, Vogeley, Weinberg, \&
  Yanny}]{eisenstein_spectroscopic_2001}
Eisenstein, D.~J., {et~al.} 2001, Astronomical Journal, 122, 2267

\bibitem[{Eisenstein {et~al.}(2011)Eisenstein, Weinberg, Agol, Aihara,
  Allende~Prieto, Anderson, Arns, Aubourg, Bailey, Balbinot, Barkhouser, Beers,
  Berlind, Bickerton, Bizyaev, Blanton, Bochanski, Bolton, Bosman, Bovy,
  Brewington, Brandt, Breslauer, Brinkmann, Brown, Brownstein, Burger, Busca,
  Campbell, Cargile, Carithers, Carlberg, Carr, Chen, Chiappini, Comparat,
  Connolly, Cortes, Croft, da~Costa, Cunha, Davenport, Dawson, De~Lee, Porto~de
  Mello, de~Simoni, Dean, Dhital, Ealet, Ebelke, Edmondson, Eiting, Escoffier,
  Esposito, Evans, Fan, Femenia~Castella, Dutra~Ferreira, Fitzgerald, Fleming,
  {Font-Ribera}, Ford, Gaudi, Ge, Ghezzi, Gillespie, Gilmore, Girardi, Gott,
  Gould, Grebel, Gunn, Harding, Harris, Hawley, Hearty, Ho, Hogg, Holtzman,
  Honscheid, Inada, Ivans, Jiang, Jiang, Johnson, Jordan, Jordan, Kazin,
  Kirkby, Klaene, Kneib, Knapp, Kochanek, Koesterke, Kollmeier, Kron, Lang,
  Le~Goff, Lee, Lee, Lin, Liu, Long, Loomis, Lucatello, Lundgren, Lupton, Ma,
  Ma, Mack, Maia, Majewski, Makler, Mandelbaum, Maraston, Margala, Maseman,
  Masters, {McDonald}, {McGreer}, {McMahon}, Mena~Requejo, Menard,
  {Miralda-Escude}, Morrison, Mullally, Muna, Murayama, Myers, Naugle,
  Fausti~Neto, Cuong~Nguyen, Nichol, Nidever, {O'Connell}, Ogando, Olmstead,
  Oravetz, Padmanabhan, Paegert, {Palanque-Delabrouille}, Pan, Pandey, Parejko,
  Paris, Pellegrini, Pepper, Percival, Petitjean, Pfaffenberger, Pforr, Phleps,
  Pichon, Pieri, Prada, {Price-Whelan}, Raddick, Ramos, Ryle, Reid, Rich,
  Richards, Rieke, Rieke, Rix, Robin, {Rocha-Pinto}, Rockosi, Roe, Rollinde,
  Ross, Ross, Rossetto, Sanchez, Santiago, Sayres, Schiavon, Schlegel,
  Schlesinger, Schmidt, Schneider, Sellgren, Shelden, Sheldon, Shetrone, Shu,
  Silverman, Simmerer, Simmons, Sivarani, Skrutskie, Slosar, Smee, Smith,
  Snedden, Stassun, Steele, Steinmetz, Stockett, Stollberg, Strauss, Tanaka,
  Thakar, Thomas, Tinker, Tofflemire, Tojeiro, Tremonti, Vargas~Magana, Verde,
  Vogt, Wake, Wan, Wang, Weaver, White, White, Wilson, Wisniewski,
  {Wood-Vasey}, Yanny, Yasuda, Yeche, York, Young, Zasowski, Zehavi, \&
  Zhao}]{eisenstein_sdss-iii:_2011}
---. 2011, {SDSS-III:} Massive Spectroscopic Surveys of the Distant Universe,
  the Milky Way Galaxy, and {Extra-Solar} Planetary Systems

\bibitem[{Faber(1973)}]{faber_variations_1973}
Faber, S.~M. 1973, The Astrophysical Journal, 179, 731

\bibitem[{Fairley {et~al.}(2002)Fairley, Jones, Wake, Collins, Burke, Nichol,
  \& Romer}]{fairley_galaxy_2002}
Fairley, B.~W., Jones, L.~R., Wake, D.~A., Collins, C.~A., Burke, D.~J.,
  Nichol, R.~C., \& Romer, A.~K. 2002, Monthly Notices of the Royal
  Astronomical Society, 330, 755

\bibitem[{Fukugita {et~al.}(1996)Fukugita, Ichikawa, Gunn, Doi, Shimasaku, \&
  Schneider}]{fukugita_sloan_1996}
Fukugita, M., Ichikawa, T., Gunn, J.~E., Doi, M., Shimasaku, K., \& Schneider,
  D.~P. 1996, The Astronomical Journal, 111, 1748

\bibitem[{Gaidos(1997)}]{gaidos_galaxy_1997}
Gaidos, E.~J. 1997, The Astronomical Journal, 113, 117

\bibitem[{Geller \& Huchra(1983)}]{geller_groups_1983}
Geller, M.~J., \& Huchra, J.~P. 1983, The Astrophysical Journal Supplement
  Series, 52, 61

\bibitem[{Goto {et~al.}(2003)Goto, Okamura, Yagi, Sheth, Bahcall, Zabel,
  Crouch, Sekiguchi, Annis, Bernardi, Chong, Gómez, Hansen, Kim, Knudson,
  {McKay}, \& Miller}]{goto_morphological_2003}
Goto, T., {et~al.} 2003, Publications of the Astronomical Society of Japan, 55,
  739

\bibitem[{Gunn {et~al.}(1998)Gunn, Carr, Rockosi, Sekiguchi, Berry, Elms,
  de~Haas, Ivezić, Knapp, Lupton, Pauls, Simcoe, Hirsch, Sanford, Wang, York,
  Harris, Annis, Bartozek, Boroski, Bakken, Haldeman, Kent, Holm, Holmgren,
  Petravick, Prosapio, Rechenmacher, Doi, Fukugita, Shimasaku, Okada, Hull,
  Siegmund, Mannery, Blouke, Heidtman, Schneider, Lucinio, \&
  Brinkman}]{gunn_sloan_1998}
Gunn, J.~E., {et~al.} 1998, The Astronomical Journal, 116, 3040

\bibitem[{Gunn {et~al.}(2006)Gunn, Siegmund, Mannery, Owen, Hull, Leger, Carey,
  Knapp, York, Boroski, Kent, Lupton, Rockosi, Evans, Waddell, Anderson, Annis,
  Barentine, Bartoszek, Bastian, Bracker, Brewington, Briegel, Brinkmann,
  Brown, Carr, Czarapata, Drennan, Dombeck, Federwitz, Gillespie, Gonzales,
  Hansen, Harvanek, Hayes, Jordan, Kinney, Klaene, Kleinman, Kron, Kresinski,
  Lee, Limmongkol, Lindenmeyer, Long, Loomis, {McGehee}, Mantsch, Neilsen,
  Neswold, Newman, Nitta, Peoples, Pier, Prieto, Prosapio, Rivetta, Schneider,
  Snedden, \& Wang}]{gunn_2.5_2006}
---. 2006, The Astronomical Journal, 131, 2332

\bibitem[{Hill {et~al.}(2011)Hill, Kelvin, Driver, Robotham, Cameron, Cross,
  Andrae, Baldry, Bamford, {Bland-Hawthorn}, Brough, Conselice, Dye, Hopkins,
  Liske, Loveday, Norberg, Peacock, Croom, Frenk, Graham, Jones, Kuijken,
  Madore, Nichol, Parkinson, Phillipps, Pimbblet, Popescu, Prescott, Seibert,
  Sharp, Sutherland, Thomas, Tuffs, \& van Kampen}]{hill_galaxy_2011}
Hill, D.~T., {et~al.} 2011, Monthly Notices of the Royal Astronomical Society,
  412, 765

\bibitem[{Hogg {et~al.}(2004)Hogg, Blanton, Brinchmann, Eisenstein, Schlegel,
  Gunn, {McKay}, Rix, Bahcall, Brinkmann, \& Meiksin}]{hogg_dependence_2004}
Hogg, D.~W., {et~al.} 2004, The Astrophysical Journal, 601, L29

\bibitem[{Huchra \& Geller(1982)}]{huchra_groups_1982}
Huchra, J.~P., \& Geller, M.~J. 1982, The Astrophysical Journal, 257, 423

\bibitem[{Jørgensen(1999)}]{jrgensen_e_1999}
Jørgensen, I. 1999, Monthly Notices of the Royal Astronomical Society, 306,
  607

\bibitem[{Kauffmann {et~al.}(2003)Kauffmann, Heckman, White, Charlot, Tremonti,
  Brinchmann, Bruzual, Peng, Seibert, Bernardi, Blanton, Brinkmann, Castander,
  Csábai, Fukugita, Ivezic, Munn, Nichol, Padmanabhan, Thakar, Weinberg, \&
  York}]{kauffmann_stellar_2003}
Kauffmann, G., {et~al.} 2003, Monthly Notices of the Royal Astronomical
  Society, 341, 33

\bibitem[{Kaviraj {et~al.}(2009)Kaviraj, Peirani, Khochfar, Silk, \&
  Kay}]{kaviraj_role_2009}
Kaviraj, S., Peirani, S., Khochfar, S., Silk, J., \& Kay, S. 2009, Monthly
  Notices of the Royal Astronomical Society, 394, 1713

\bibitem[{Kaviraj {et~al.}(2008)Kaviraj, Khochfar, Schawinski, Yi, Gawiser,
  Silk, Virani, Cardamone, van Dokkum, \& Urry}]{kaviraj_uv_2008}
Kaviraj, S., {et~al.} 2008, Monthly Notices of the Royal Astronomical Society,
  388, 67

\bibitem[{Khosroshahi {et~al.}(2004)Khosroshahi, Jones, \&
  Ponman}]{khosroshahi_old_2004}
Khosroshahi, H.~G., Jones, L.~R., \& Ponman, T.~J. 2004, Monthly Notices of the
  Royal Astronomical Society, 349, 1240

\bibitem[{Kormendy \& Djorgovski(1989)}]{kormendy_surface_1989}
Kormendy, J., \& Djorgovski, S. 1989, Annual Review of Astronomy and
  Astrophysics, 27, 235

\bibitem[{Labbé {et~al.}(2003)Labbé, Franx, Rudnick, Schreiber, Rix,
  Moorwood, van Dokkum, van~der Werf, Röttgering, van Starkenburg, van~der
  Wel, Kuijken, \& Daddi}]{labbe_ultradeep_2003}
Labbé, I., {et~al.} 2003, The Astronomical Journal, 125, 1107

\bibitem[{Lin {et~al.}(2004)Lin, Mohr, \& Stanford}]{lin_k-band_2004}
Lin, Y., Mohr, J.~J., \& Stanford, S.~A. 2004, The Astrophysical Journal, 610,
  745

\bibitem[{Lobo {et~al.}(1997)Lobo, Biviano, Durret, Gerbal, Le~Fevre, Mazure,
  \& Slezak}]{lobo_environmental_1997}
Lobo, C., Biviano, A., Durret, F., Gerbal, D., Le~Fevre, O., Mazure, A., \&
  Slezak, E. 1997, Astronomy and Astrophysics, 317, 385

\bibitem[{Loh {et~al.}(2008)Loh, Ellingson, Yee, Gilbank, Gladders, \&
  Barrientos}]{loh_color_2008}
Loh, Y., Ellingson, E., Yee, H. K.~C., Gilbank, D.~G., Gladders, M.~D., \&
  Barrientos, L.~F. 2008, The Astrophysical Journal, 680, 214

\bibitem[{Lu {et~al.}(2009)Lu, Gilbank, Balogh, \& Bognat}]{lu_recent_2009}
Lu, T., Gilbank, D.~G., Balogh, M.~L., \& Bognat, A. 2009, Monthly Notices of
  the Royal Astronomical Society, 399, 1858

\bibitem[{Lumsden {et~al.}(1997)Lumsden, Collins, Nichol, Eke, \&
  Guzzo}]{lumsden_edinburgh-durham_1997}
Lumsden, S.~L., Collins, C.~A., Nichol, R.~C., Eke, V.~R., \& Guzzo, L. 1997,
  Monthly Notices of the Royal Astronomical Society, 290, 119

\bibitem[{Maraston {et~al.}(2009)Maraston, Strömbäck, Thomas, Wake, \&
  Nichol}]{maraston_modelling_2009}
Maraston, C., Strömbäck, G., Thomas, D., Wake, D.~A., \& Nichol, R.~C. 2009,
  Monthly Notices of the Royal Astronomical Society, 394, L107

\bibitem[{Markwardt(2009)}]{markwardt_non-linear_2009}
Markwardt, C.~B. 2009, in Astronomical Data Analysis Software and Systems
  {XVIII}, Vol. 411, 251

\bibitem[{Martini {et~al.}(2006)Martini, Kelson, Kim, Mulchaey, \&
  Athey}]{martini_spectroscopic_2006}
Martini, P., Kelson, D.~D., Kim, E., Mulchaey, J.~S., \& Athey, A.~A. 2006, The
  Astrophysical Journal, 644, 116

\bibitem[{Masjedi {et~al.}(2008)Masjedi, Hogg, \&
  Blanton}]{masjedi_growth_2008}
Masjedi, M., Hogg, D.~W., \& Blanton, M.~R. 2008, The Astrophysical Journal,
  679, 260

\bibitem[{Masjedi {et~al.}(2006)Masjedi, Hogg, Cool, Eisenstein, Blanton,
  Zehavi, Berlind, Bell, Schneider, Warren, \& Brinkmann}]{masjedi_very_2006}
Masjedi, M., {et~al.} 2006, The Astrophysical Journal, 644, 54

\bibitem[{Masters {et~al.}(2011)Masters, Maraston, Nichol, Thomas, Beifiori,
  Bundy, Edmondson, Higgs, Leauthaud, Mandelbaum, Pforr, Ross, Ross, Schneider,
  Skibba, Tinker, Tojeiro, Wake, Brinkmann, \&
  Weaver}]{masters_morphology_2011}
Masters, K.~L., {et~al.} 2011, Monthly Notices of the Royal Astronomical
  Society, no

\bibitem[{{McIntosh} {et~al.}(2008){McIntosh}, Guo, Hertzberg, Katz, Mo,
  van~den Bosch, \& Yang}]{mcintosh_ongoing_2008}
{McIntosh}, D.~H., Guo, Y., Hertzberg, J., Katz, N., Mo, H.~J., van~den Bosch,
  F.~C., \& Yang, X. 2008, Monthly Notices of the Royal Astronomical Society,
  388, 1537

\bibitem[{Milosavljević {et~al.}(2006)Milosavljević, Miller, Furlanetto, \&
  Cooray}]{milosavljevi_cluster_2006}
Milosavljević, M., Miller, C.~J., Furlanetto, S.~R., \& Cooray, A. 2006, The
  Astrophysical Journal, 637, L9

\bibitem[{Muzzin {et~al.}(2009)Muzzin, Wilson, Yee, Hoekstra, Gilbank, Surace,
  Lacy, Blindert, Majumdar, Demarco, Gardner, Gladders, \&
  Lonsdale}]{muzzin_spectroscopic_2009}
Muzzin, A., {et~al.} 2009, The Astrophysical Journal, 698, 1934

\bibitem[{Naab {et~al.}(2009)Naab, Johansson, \& Ostriker}]{naab_minor_2009}
Naab, T., Johansson, P.~H., \& Ostriker, J.~P. 2009, The Astrophysical Journal,
  699, L178

\bibitem[{Naab {et~al.}(2007)Naab, Johansson, Ostriker, \&
  Efstathiou}]{naab_formation_2007}
Naab, T., Johansson, P.~H., Ostriker, J.~P., \& Efstathiou, G. 2007,
  Astrophysical Journal, 658, 710

\bibitem[{Naab {et~al.}(2006)Naab, Khochfar, \& Burkert}]{naab_properties_2006}
Naab, T., Khochfar, S., \& Burkert, A. 2006, The Astrophysical Journal, 636,
  L81

\bibitem[{Patton {et~al.}(2002)Patton, Pritchet, Carlberg, Marzke, Yee, Hall,
  Lin, Morris, Sawicki, Shepherd, \& Wirth}]{patton_dynamically_2002}
Patton, D.~R., {et~al.} 2002, The Astrophysical Journal, 565, 208

\bibitem[{Peletier(1989)}]{peletier_elliptical_1989}
Peletier, R.~F. 1989, Elliptical Galaxies - Structure and Stellar Content,
  {http://adsabs.harvard.edu/abs/1989PhDT.......149P}

\bibitem[{Pier {et~al.}(2003)Pier, Munn, Hindsley, Hennessy, Kent, Lupton, \&
  Ivezić}]{pier_astrometric_2003}
Pier, J.~R., Munn, J.~A., Hindsley, R.~B., Hennessy, G.~S., Kent, S.~M.,
  Lupton, R.~H., \& Ivezić, Å. 2003, The Astronomical Journal, 125, 1559

\bibitem[{Ponman {et~al.}(1994)Ponman, Allan, Jones, Merrifield, {McHardy},
  Lehto, \& Luppino}]{ponman_possible_1994}
Ponman, T.~J., Allan, D.~J., Jones, L.~R., Merrifield, M., {McHardy}, I.~M.,
  Lehto, H.~J., \& Luppino, G.~A. 1994, Nature, 369, 462

\bibitem[{Ramella {et~al.}(2000)Ramella, Biviano, Boschin, Bardelli, Scodeggio,
  Borgani, Benoist, da~Costa, Girardi, Nonino, \&
  Olsen}]{ramella_spectroscopic_2000}
Ramella, M., {et~al.} 2000, Astronomy and Astrophysics, 360, 861

\bibitem[{Ramos~Almeida {et~al.}(2011)Ramos~Almeida, Tadhunter, Inskip,
  Morganti, Holt, \& Dicken}]{ramos_almeida_optical_2011}
Ramos~Almeida, C., Tadhunter, C.~N., Inskip, K.~J., Morganti, R., Holt, J., \&
  Dicken, D. 2011, Monthly Notices of the Royal Astronomical Society, 410, 1550

\bibitem[{Reid \& Spergel(2009)}]{reid_constraining_2009}
Reid, B.~A., \& Spergel, D.~N. 2009, The Astrophysical Journal, 698, 143

\bibitem[{Sandage \& Hardy(1973)}]{sandage_redshift-distance_1973}
Sandage, A., \& Hardy, E. 1973, The Astrophysical Journal, 183, 743

\bibitem[{Schechter(1976)}]{schechter_analytic_1976}
Schechter, P. 1976, The Astrophysical Journal, 203, 297

\bibitem[{Schlegel {et~al.}(2009)Schlegel, White, \&
  Eisenstein}]{schlegel_baryon_2009}
Schlegel, D., White, M., \& Eisenstein, D. 2009, in astro2010: The Astronomy
  and Astrophysics Decadal Survey, Vol. 2010, 314

\bibitem[{Schneider {et~al.}(1983)Schneider, Gunn, \&
  Hoessel}]{schneider_ccd_1983}
Schneider, D.~P., Gunn, J.~E., \& Hoessel, J.~G. 1983, The Astrophysical
  Journal, 268, 476

\bibitem[{Schweizer \& Seitzer(1992)}]{schweizer_correlations_1992}
Schweizer, F., \& Seitzer, P. 1992, Astronomical Journal, 104, 1039

\bibitem[{Smith {et~al.}(2002)Smith, Tucker, Kent, Richmond, Fukugita,
  Ichikawa, Ichikawa, Jorgensen, Uomoto, Gunn, Hamabe, Watanabe, Tolea, Henden,
  Annis, Pier, {McKay}, Brinkmann, Chen, Holtzman, Shimasaku, \&
  York}]{smith_ugriz_2002}
Smith, J.~A., {et~al.} 2002, The Astronomical Journal, 123, 2121

\bibitem[{Stoughton {et~al.}(2002)Stoughton, Lupton, Bernardi, Blanton, Burles,
  Castander, Connolly, Eisenstein, Frieman, Hennessy, Hindsley, Ivezić, Kent,
  Kunszt, Lee, Meiksin, Munn, Newberg, Nichol, Nicinski, Pier, Richards,
  Richmond, Schlegel, Smith, Strauss, {SubbaRao}, Szalay, Thakar, Tucker,
  Vanden~Berk, Yanny, Adelman, Anderson, Anderson, Annis, Bahcall, Bakken,
  Bartelmann, Bastian, Bauer, Berman, Böhringer, Boroski, Bracker, Briegel,
  Briggs, Brinkmann, Brunner, Carey, Carr, Chen, Christian, Colestock, Crocker,
  Csabai, Czarapata, Dalcanton, Davidsen, Davis, Dehnen, Dodelson, Doi,
  Dombeck, Donahue, Ellman, Elms, Evans, Eyer, Fan, Federwitz, Friedman,
  Fukugita, Gal, Gillespie, Glazebrook, Gray, Grebel, Greenawalt, Greene, Gunn,
  de~Haas, Haiman, Haldeman, Hall, Hamabe, Hansen, Harris, Harris, Harvanek,
  Hawley, Hayes, Heckman, Helmi, Henden, Hogan, Hogg, Holmgren, Holtzman,
  Huang, Hull, Ichikawa, Ichikawa, Johnston, Kauffmann, Kim, Kimball, Kinney,
  Klaene, Kleinman, Klypin, Knapp, Korienek, Krolik, Kron, Krzesiński, Lamb,
  Leger, Limmongkol, Lindenmeyer, Long, Loomis, Loveday, {MacKinnon}, Mannery,
  Mantsch, Margon, {McGehee}, {McKay}, {McLean}, Menou, Merelli, Mo, Monet,
  Nakamura, Narayanan, Nash, Neilsen, Newman, Nitta, Odenkirchen, Okada,
  Okamura, Ostriker, Owen, Pauls, Peoples, Peterson, Petravick, Pope, Pordes,
  Postman, Prosapio, Quinn, Rechenmacher, Rivetta, Rix, Rockosi, Rosner,
  Ruthmansdorfer, Sandford, Schneider, Scranton, Sekiguchi, Sergey, Sheth,
  Shimasaku, Smee, Snedden, Stebbins, Stubbs, Szapudi, Szkody, Szokoly,
  Tabachnik, Tsvetanov, Uomoto, Vogeley, Voges, Waddell, Walterbos, Wang,
  Watanabe, Weinberg, White, White, Wilhite, Wolfe, Yasuda, York, Zehavi, \&
  Zheng}]{stoughton_sloan_2002}
Stoughton, C., {et~al.} 2002, The Astronomical Journal, 123, 485

\bibitem[{Tal \& van Dokkum(2011)}]{tal_faint_2011}
Tal, T., \& van Dokkum, P.~G. 2011, The Astrophysical Journal, 731, 89

\bibitem[{Tal {et~al.}(2009)Tal, van Dokkum, Nelan, \&
  Bezanson}]{tal_frequency_2009}
Tal, T., van Dokkum, P.~G., Nelan, J., \& Bezanson, R. 2009, The Astronomical
  Journal, 138, 1417

\bibitem[{Tanaka {et~al.}(2010)Tanaka, Finoguenov, \&
  Ueda}]{tanaka_spectroscopically_2010}
Tanaka, M., Finoguenov, A., \& Ueda, Y. 2010, The Astrophysical Journal, 716,
  L152

\bibitem[{Taylor {et~al.}(2011)Taylor, Hopkins, Baldry, Brown, Driver, Kelvin,
  Hill, Robotham, {Bland-Hawthorn}, Jones, Sharp, Thomas, Liske, Loveday,
  Norberg, Peacock, Bamford, Brough, Colless, Cameron, Conselice, Croom, Frenk,
  Gunawardhana, Kuijken, Nichol, Parkinson, Phillipps, Pimbblet, Popescu,
  Prescott, Sutherland, Tuffs, van Kampen, \& Wijesinghe}]{taylor_galaxy_2011}
Taylor, E.~N., {et~al.} 2011, Galaxy And Mass Assembly: Stellar Mass Estimates

\bibitem[{Thomas {et~al.}(2005)Thomas, Maraston, Bender, \& Mendes~de
  Oliveira}]{thomas_epochs_2005}
Thomas, D., Maraston, C., Bender, R., \& Mendes~de Oliveira, C. 2005, The
  Astrophysical Journal, 621, 673

\bibitem[{Tojeiro \& Percival(2010)}]{tojeiro_evolution_2010}
Tojeiro, R., \& Percival, W.~J. 2010, Monthly Notices of the Royal Astronomical
  Society, 405, 2534

\bibitem[{Tojeiro {et~al.}(2011)Tojeiro, Percival, Heavens, \&
  Jimenez}]{tojeiro_stellar_2011}
Tojeiro, R., Percival, W.~J., Heavens, A.~F., \& Jimenez, R. 2011, Monthly
  Notices of the Royal Astronomical Society, 413, 434

\bibitem[{Trager {et~al.}(2000)Trager, Faber, Worthey, \&
  González}]{trager_stellar_2000}
Trager, S.~C., Faber, S.~M., Worthey, G., \& González, J.~J. 2000, The
  Astronomical Journal, 119, 1645

\bibitem[{Tran {et~al.}(2005)Tran, van Dokkum, Franx, Illingworth, Kelson, \&
  Schreiber}]{tran_spectroscopic_2005}
Tran, K.~H., van Dokkum, P., Franx, M., Illingworth, G.~D., Kelson, D.~D., \&
  Schreiber, N. M.~F. 2005, The Astrophysical Journal, 627, L25

\bibitem[{Tremaine \& Richstone(1977)}]{tremaine_test_1977}
Tremaine, S.~D., \& Richstone, D.~O. 1977, The Astrophysical Journal, 212, 311

\bibitem[{Valotto {et~al.}(1997)Valotto, Nicotra, Muriel, \&
  Lambas}]{valotto_luminosity_1997}
Valotto, C.~A., Nicotra, M.~A., Muriel, H., \& Lambas, D.~G. 1997, The
  Astrophysical Journal, 479, 90

\bibitem[{van~den Bosch {et~al.}(2008)van~den Bosch, Aquino, Yang, Mo,
  Pasquali, {McIntosh}, Weinmann, \& Kang}]{van_den_bosch_importance_2008}
van~den Bosch, F.~C., Aquino, D., Yang, X., Mo, H.~J., Pasquali, A.,
  {McIntosh}, D.~H., Weinmann, S.~M., \& Kang, X. 2008, Monthly Notices of the
  Royal Astronomical Society, 387, 79

\bibitem[{van~den Bosch {et~al.}(2007)van~den Bosch, Yang, Mo, Weinmann,
  Macciò, More, Cacciato, Skibba, \& Kang}]{van_den_bosch_towards_2007}
van~den Bosch, F.~C., {et~al.} 2007, Monthly Notices of the Royal Astronomical
  Society, 376, 841

\bibitem[{van Dokkum(2005)}]{van_dokkum_recent_2005}
van Dokkum, P.~G. 2005, The Astronomical Journal, 130, 2647

\bibitem[{van Dokkum \& Franx(1996)}]{van_dokkum_fundamental_1996}
van Dokkum, P.~G., \& Franx, M. 1996, Monthly Notices of the Royal Astronomical
  Society, 281, 985

\bibitem[{van Dokkum {et~al.}(1999)van Dokkum, Franx, Fabricant, Kelson, \&
  Illingworth}]{van_dokkum_high_1999}
van Dokkum, P.~G., Franx, M., Fabricant, D., Kelson, D.~D., \& Illingworth,
  G.~D. 1999, The Astrophysical Journal, 520, L95

\bibitem[{van Dokkum {et~al.}(2010)van Dokkum, Whitaker, Brammer, Franx, Kriek,
  Labbé, Marchesini, Quadri, Bezanson, Illingworth, Muzzin, Rudnick, Tal, \&
  Wake}]{van_dokkum_growth_2010}
van Dokkum, P.~G., {et~al.} 2010, The Astrophysical Journal, 709, 1018

\bibitem[{Wake {et~al.}(2005)Wake, Collins, Nichol, Jones, \&
  Burke}]{wake_environmental_2005}
Wake, D.~A., Collins, C.~A., Nichol, R.~C., Jones, L.~R., \& Burke, D.~J. 2005,
  The Astrophysical Journal, 627, 186

\bibitem[{Wake {et~al.}(2006)Wake, Nichol, Eisenstein, Loveday, Edge, Cannon,
  Smail, Schneider, Scranton, Carson, Ross, Brunner, Colless, Couch, Croom,
  Driver, DaÂngela, Jester, De~Propris, Drinkwater, {Bland‐Hawthorn},
  Pimbblet, Roseboom, Shanks, Sharp, \& Brinkmann}]{wake_2df_2006}
Wake, D.~A., {et~al.} 2006, Monthly Notices of the Royal Astronomical Society,
  372, 537

\bibitem[{Wake {et~al.}(2008)Wake, Sheth, Nichol, Baugh, {Bland-Hawthorn},
  Colless, Couch, Croom, de~Propris, Drinkwater, Edge, Loveday, Lam, Pimbblet,
  Roseboom, Ross, Schneider, Shanks, \& Sharp}]{wake_2df-sdss_2008}
---. 2008, Monthly Notices of the Royal Astronomical Society, 387, 1045

\bibitem[{Weinmann {et~al.}(2009)Weinmann, Kauffmann, van~den Bosch, Pasquali,
  {McIntosh}, Mo, Yang, \& Guo}]{weinmann_environmental_2009}
Weinmann, S.~M., Kauffmann, G., van~den Bosch, F.~C., Pasquali, A., {McIntosh},
  D.~H., Mo, H., Yang, X., \& Guo, Y. 2009, Monthly Notices of the Royal
  Astronomical Society, 394, 1213

\bibitem[{White {et~al.}(2011)White, Blanton, Bolton, Schlegel, Tinker,
  Berlind, da~Costa, Kazin, Lin, Maia, {McBride}, Padmanabhan, Parejko,
  Percival, Prada, Ramos, Sheldon, de~Simoni, Skibba, Thomas, Wake, Zehavi,
  Zheng, Nichol, Schneider, Strauss, Weaver, \&
  Weinberg}]{white_clustering_2011}
White, M., {et~al.} 2011, The Astrophysical Journal, 728, 126

\bibitem[{Wilson {et~al.}(1997)Wilson, Smail, Ellis, \&
  Couch}]{wilson_faint_1997}
Wilson, G., Smail, I., Ellis, R.~S., \& Couch, W.~J. 1997, Monthly Notices of
  the Royal Astronomical Society, 284, 915

\bibitem[{Wilson {et~al.}(2009)Wilson, Muzzin, Yee, Lacy, Surace, Gilbank,
  Blindert, Hoekstra, Majumdar, Demarco, Gardner, Gladders, \&
  Lonsdale}]{wilson_spectroscopic_2009}
Wilson, G., {et~al.} 2009, The Astrophysical Journal, 698, 1943

\bibitem[{Worthey {et~al.}(1992)Worthey, Faber, \& Gonzalez}]{worthey_mg_1992}
Worthey, G., Faber, S.~M., \& Gonzalez, J.~J. 1992, The Astrophysical Journal,
  398, 69

\bibitem[{Yang {et~al.}(2008)Yang, Mo, \& van~den Bosch}]{yang_galaxy_2008}
Yang, X., Mo, H.~J., \& van~den Bosch, F.~C. 2008, The Astrophysical Journal,
  676, 248

\bibitem[{Yang {et~al.}(2005)Yang, Mo, van~den Bosch, \&
  Jing}]{yang_halo-based_2005}
Yang, X., Mo, H.~J., van~den Bosch, F.~C., \& Jing, Y.~P. 2005, Monthly Notices
  of the Royal Astronomical Society, 356, 1293

\bibitem[{York {et~al.}(2000)York, Adelman, Anderson, Anderson, Annis, Bahcall,
  Bakken, Barkhouser, Bastian, Berman, Boroski, Bracker, Briegel, Briggs,
  Brinkmann, Brunner, Burles, Carey, Carr, Castander, Chen, Colestock,
  Connolly, Crocker, Csabai, Czarapata, Davis, Doi, Dombeck, Eisenstein,
  Ellman, Elms, Evans, Fan, Federwitz, Fiscelli, Friedman, Frieman, Fukugita,
  Gillespie, Gunn, Gurbani, de~Haas, Haldeman, Harris, Hayes, Heckman,
  Hennessy, Hindsley, Holm, Holmgren, Huang, Hull, Husby, Ichikawa, Ichikawa,
  Ivezić, Kent, Kim, Kinney, Klaene, Kleinman, Kleinman, Knapp, Korienek,
  Kron, Kunszt, Lamb, Lee, Leger, Limmongkol, Lindenmeyer, Long, Loomis,
  Loveday, Lucinio, Lupton, {MacKinnon}, Mannery, Mantsch, Margon, {McGehee},
  {McKay}, Meiksin, Merelli, Monet, Munn, Narayanan, Nash, Neilsen, Neswold,
  Newberg, Nichol, Nicinski, Nonino, Okada, Okamura, Ostriker, Owen, Pauls,
  Peoples, Peterson, Petravick, Pier, Pope, Pordes, Prosapio, Rechenmacher,
  Quinn, Richards, Richmond, Rivetta, Rockosi, Ruthmansdorfer, Sandford,
  Schlegel, Schneider, Sekiguchi, Sergey, Shimasaku, Siegmund, Smee, Smith,
  Snedden, Stone, Stoughton, Strauss, Stubbs, {SubbaRao}, Szalay, Szapudi,
  Szokoly, Thakar, Tremonti, Tucker, Uomoto, Vanden~Berk, Vogeley, Waddell,
  Wang, Watanabe, Weinberg, Yanny, \& Yasuda}]{york_sloan_2000}
York, D.~G., {et~al.} 2000, The Astronomical Journal, 120, 1579

\bibitem[{Zheng {et~al.}(2009)Zheng, Zehavi, Eisenstein, Weinberg, \&
  Jing}]{zheng_halo_2009}
Zheng, Z., Zehavi, I., Eisenstein, D.~J., Weinberg, D.~H., \& Jing, Y.~P. 2009,
  The Astrophysical Journal, 707, 554

\end{thebibliography}
\bibliographystyle{apj}

\appendix
 \section{Comparison between SDSS and SExtractor magnitudes}
  Aperture based photometry can lead to a systematic misestimate of
  total flux measurements compared to model based photometry.
  \cite{taylor_galaxy_2011} performed an extensive comparison of
  several flux estimates and showed that SExtractor \emph{AUTO}
  magnitudes are up to 10\% fainter than SDSS \emph{model}
  magnitudes.
  \cite{taylor_galaxy_2011} further found that this value is dependent
  on the light profile slope of the best-fit galaxy model.

\begin{figure}[h]
    \center
    \includegraphics[width=0.47\textwidth]{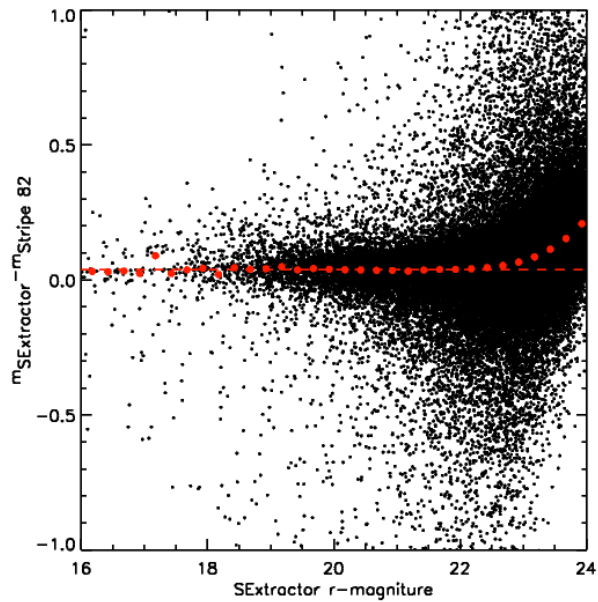}
    \caption{Difference between SExtractor \emph{AUTO} magnitudes and
      SDSS \emph{model} magnitudes for sources in 150 SDSS Stripe
      82fields.
      The black data points represent the measured values of all
      sources while the red thick points are the running median of the
      sample. 
      The dashed red line is a linear fit to the running median and it
      shows that SExtractor underestimates the total object flux
      compared to Stripe 82 by roughly 0.04 magnitudes.
      Assuming that SDSS \emph{model} magnitudes are a better
      assessment of the total object flux we use this fit to correct
      our SExtractor measurements.}
    \hfill
    \label{fig:secomp}
  \end{figure}

  In order to correct for this effect we extracted photometry for all
  objects in 150 SDSS fields in Stripe 82 and compared the resulting
  \emph{AUTO} magnitudes to \emph{model} magnitudes from the Stripe 82
  catalogs.
  This comparison is shown in figure \ref{fig:secomp}, where the
  difference between the measurements is plotted against the
  SExtractor values.
  We then fitted a line to the running median of this distribution and
  derived a correction relation for the extracted magnitudes used
  throughout this paper, assuming that Stripe 82 \emph{model}
  magnitudes better reflect the total flux of the studied galaxies.
  The correction factor is essentially constant and it has a value of
  roughly 0.04 magnitudes between $m_{SE}=17$ and $m_{SE}=22$.
  The distribution plotted in Figure \ref{fig:secomp} is consistent
  with the relation between SDSS \emph{model} and SExtractor
  \emph{AUTO} measurements found by \cite{taylor_galaxy_2011} for
  objects brighter than roughly 17.5 magnitudes.
  \cite{hill_galaxy_2011} compared SDSS and SExtractor flux values over
  a broad range of magnitudes and also found a small offset between the
  measurements.

  We note that since all galaxies in our sample do not have the same
  light profile slope, we introduce a error of up to 10\% to the
  measured luminosities by applying a single correction relation for
  all galaxies.
  This error could potentially translate to a systematic offset in the
  flux measurement of the LRGs themselves if the LRG light profile is
  significantly steeper than the typical satellite profile.
  However, since the absolute fraction of missed light at any given
  magnitude is small, we expect this contribution to be insignificant
  compared to other sources of scatter in this study.
  
\end{document}